\documentclass[journal]{IEEEtran}
\usepackage{amsmath,amsfonts}
\usepackage{algorithmic}
\usepackage{algorithm}
\usepackage{array}
\usepackage[caption=false,font=normalsize,labelfont=sf,textfont=sf]{subfig}
\usepackage{textcomp}
\usepackage{stfloats}
\usepackage{url}
\usepackage{verbatim}
\usepackage{graphicx}
\usepackage{cite}
\usepackage{xcolor}

\usepackage[table,xcdraw]{xcolor}
\usepackage{hyperref} 

\begin{document}

\title{A Heterogeneous Neural Network Accelerator for End-to-End Multitask RF Signal Recognition}

\author{Zhifan Song, Haralampos-G. Stratigopoulos,~\IEEEmembership{Member,~IEEE}, and Hassan Aboushady,~\IEEEmembership{Senior Member,~IEEE}
\small{Sorbonne University, CNRS, LIP6, Paris, France}\vspace{-0.4cm}

\thanks{This work was funded by the Chips JU project Resilient Trust of the EU’s Horizon Europe research and innovation programme under Grant agreement N$^{\mbox{\scriptsize o}}$ 101112282.}}



\maketitle

\begin{abstract}
This paper presents a heterogeneous neural network accelerator for multi-task RF signal recognition, supporting automatic modulation recognition (AMR), hardware-Trojan covert channel (HT-CC) detection, and GNSS jamming classification. We introduce a compact attention-enhanced convolutional neural network (CNN) combined with LSDec, a learnable streaming decimator that enables adaptive temporal downsampling and flexible input lengths. The hardware architecture integrates a novel dual-pipeline, fused convolution–pooling engine with DMA-based streaming to minimize memory traffic and latency. Co-execution scheduling on the accelerator and SIMD-optimized CPU kernels reduces hardware resource usage while preserving high performance and task-level flexibility. Across three datasets, the proposed system achieves $\geq$ 99\% average accuracy above 4 dB Signal-to-Noise Ratios (SNRs) on the RadioML2018 dataset for AMR, 90\% on the HT-CC dataset, and 99.5\% on the GNSS-Jamming dataset. The accelerator sustains an end-to-end inference latency of 98 $\mu$s per frame, demonstrating its effectiveness for low-power, latency-critical multi-task spectrum-intelligence applications on embedded and edge devices.
\end{abstract}


\section{Introduction}
\begin{figure}[t]
\centerline{\includegraphics[width=\linewidth]{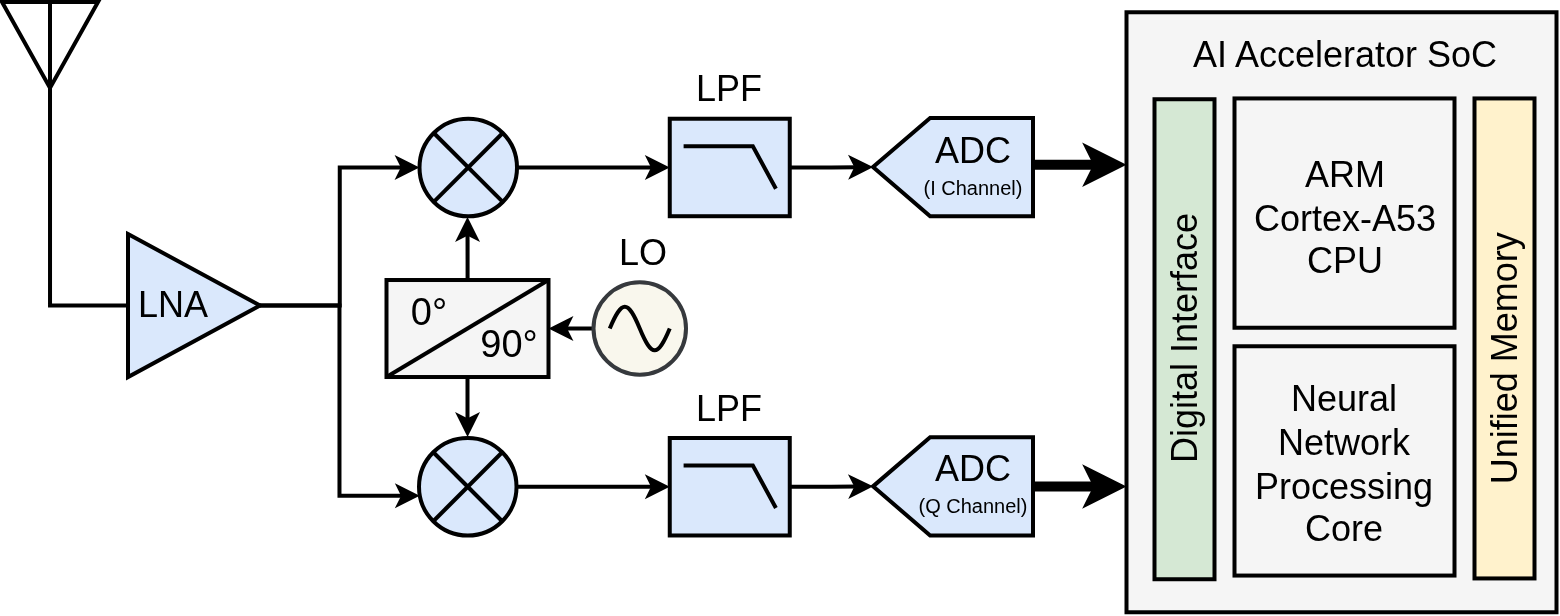}}
\caption{End-to-end RF spectrum monitoring system with dedicated AI SoC. LNA: Low-noise amplifier; LPF: Low-pass filter; LO: Local oscillator; ADC: Analog-to-digital converter.
}
\label{fig:system}
\end{figure}

\IEEEPARstart{T}{he} rapid growth of modern wireless communication systems has intensified pressure on the already congested radio spectrum, resulting in inefficient and uneven frequency band utilization~\cite{b1}. Emerging 5G networks and the large-scale deployment of Internet of Things (IoT) devices introduce dense coexistence of heterogeneous waveforms, dynamic spectrum access, and increasingly complex interference environments. Future 6G networks of ubiquitous intelligence are expected to integrate native AI-driven radio architectures and real-time spectrum cognition~\cite{aicom,airadio}, requiring embedded devices to rapidly interpret high-rate in-phase/quadrature (I/Q) samples from the RF front-end under strict latency and power constraints, as illustrated in Fig.~\ref{fig:system}.

Software-Defined Radios (SDRs)~\cite{airadio} built on FPGAs/SoCs have emerged as a flexible solution for RF sensing and edge inference, enabling real-time signal processing with tight energy budgets. Within this context, automatic modulation recognition (AMR) is a fundamental step in spectrum intelligence, enabling wireless nodes to identify signal types prior to demodulation. AMR improves spectrum traffic by removing redundant modulation metadata in the signal, supports opportunistic spectrum access for secondary users during idle intervals, and allows devices to evaluate transmission feasibility in the presence of heterogeneous technologies and interference~\cite{review}. 

Beyond AMR, modern RF environments also require robust classification of security-related phenomena, including hardware Trojan–based covert channels (HT-CCs)\cite{alan} and intentional interference such as Global Navigation Satellite System (GNSS) jamming\cite{gnss2d}. Traditional methods struggle with accuracy and computational efficiency~\cite{traditional}. Although deep learning models have demonstrated high accuracy across these tasks, most existing solutions rely on GPU evaluation or large networks with millions of parameters~\cite{lstm2amr,MCLDNN,IC-AMCNET,CGDNet,kulin}. Such approaches incur high latency and energy consumption, making them impractical for embedded SDR platforms or IoT devices. Cloud computing is likewise unsuitable due to RF data volume, privacy concerns, and stringent latency requirements~\cite{cloud}. Low-latency and power-efficient spectrum intelligence at the edge has become a critical requirement for wireless devices. 

\IEEEpubidadjcol

Recent hardware accelerators for AMR~\cite{snn,jung,Soltani,on-chip,emad,song} address parts of this challenge, but they typically focus on single-task efficiency, depend on high-end FPGAs, or have relatively high power consumption. These designs implement all layers in hardware, making them inflexible for adapting to new RF tasks or adjusting model hyperparameters without hardware redesign and resynthesis. Moreover, typical devices all have CPUs which may be underutilized.

Heterogeneous computing, where different IP cores collaboratively execute a neural network (NN), has shown strong potential for improving performance and energy efficiency in edge devices~\cite {hetero}. However, to the best of our knowledge, this paradigm has not been explored for AMR or other RF signal classification tasks, particularly on resource-constrained embedded and IoT platforms.

Many existing AMR approaches also rely on input preprocessing~\cite{constellation, spectrogram, timefrequency, dae} which not only adds latency but requires buffering full-length I/Q signals. Moreover, existing hardware accelerators are customized and mostly proposed for a single RF task, such as AMR~\cite{review2}, which requires re-design and/or re-synthesis when the task changes~\cite{MRTransformer} and often neglects the challenges of interfacing RF front-end ADCs which operate at significantly faster speeds and deploying standard NNs directly at the ADC rate leads to prohibitive power and area. Motivated by these gaps, this work introduces a heterogeneous NN accelerator, as illustrated in Fig.~\ref{fig:system}, for end-to-end multi-task RF signal recognition without preprocessing, including AMR, HT-CC detection, and GNSS jamming classification. The end-to-end system interfaces directly with an RF front-end: raw I/Q streams are fed directly into a dedicated AI System-on-Chip (SoC) for end-to-end inference. This AI accelerator leverages a heterogeneous architecture to balance computational efficiency with task-level flexibility at the edge. We extend our prior FPGA-only design in~\cite{song} with a new algorithm–architecture co-design that enables real-time edge inference on deeply embedded devices.

In this paper, we focus on system-level efficiency and flexibility.
The main contributions are as follows:
\begin{itemize}
    \item \textbf{A compact attention-enhanced integer-quantized NN architecture} designed for configurable RF signal classification, achieving high accuracy across multiple RF tasks while reducing parameter count and computational cost.

    \item \textbf{A learnable streaming decimator (LSDec)} that bridges the ADC-to-AI Core speed mismatch, enabling decimation while performing feature extraction on the fly to meet real-time constraints regardless of input length.
    
    \item \textbf{A dual-pipeline, fused convolution-maxpooling engine} with DMA-driven data transfer for further latency optimization.
    
    \item \textbf{A heterogeneous CPU -- NN accelerator framework} with an algorithm-hardware co-scheduling algorithm and task-aware flexibility, enabling adaptation of model hyperparameters on a general-purpose CPU to new RF tasks without NN accelerator hardware redesign.
    
    \item \textbf{A unified multi-task RF signal recognition system} validated on three datasets for AMR 
\cite{rml18}, HT-CC \cite{alan}, and jamming \cite{jammingdataset}, supporting processing of streaming I/Q samples at 98 $\mu$s with high classification accuracy, demonstrating strong generalizability across spectrum-intelligence and RF-security applications.
\end{itemize}

%
%
\section{Related Work}
\subsection{RF Spectrum Monitoring}

\subsubsection{Automatic Modulation Recognition}
\label{sec:review_amr}

Deep learning (DL) with NNs has demonstrated superior capability in numerous domains: Long Short-Term Memory (LSTM) networks~\cite{lstm} have shown success with time-series data but suffer from limited parallelization due to sequential structure. Transformer models~\cite{attention,vit,swin}, though powerful, are generally memory-intensive and less suitable on edge devices. Convolutional Neural Networks (CNNs) offer a better balance, providing high parallelism, reduced memory footprint, and low inference latency, making them a better fit for edge devices~\cite{mobilenetv2, mobilenetv3}. AI-based RF spectrum monitoring has become a dominant trend and models can learn discriminative representations from RF data, ensuring superior detection accuracy~\cite{review,review2}.

Conventional AMR methods fall broadly into feature-based (FB) and likelihood-based (LB) approaches. LB methods typically achieve strong accuracy at the cost of high computational cost~\cite{LB,LB2,LB3}, while FB methods rely on manual feature extraction and classifier design whose effectiveness degrades as modern communication signals grow in modulation order and variability \cite{FB,review}.

In DL-based AMR, existing models operate on raw I/Q frames or pre-processed input and transformed representations such as a constellation maps~\cite{constellation}, spectrograms~\cite{spectrogram}, time-frequency diagrams~\cite{timefrequency}, amplitude/phase signals~\cite{dae}, etc. The first seminal work is a 4-layer CNN model~\cite{rml16} which demonstrated clear advantages over traditional ML techniques trees and SVM, but it was designed for 128-length input while the model parameters scale significantly for longer I/Q sequences, and it suffers from limited accuracy~\cite{review2} on more challenging over-the-air signals~\cite{rml18}. 

Recurrent architectures excel in time-series processing and have also been explored. An LSTM~\cite{lstm2amr} with amplitude/phase-transformed input yields high accuracy but incurs substantial computational cost. A lightweight LSTM autoencoder~\cite{dae} was proposed and targeted resource-constrained platforms, reducing parameter count, yet recurrent dependencies still hinder parallelism and require full input buffering, resulting in intrinsically increased latency. CNN with Gaussian noise \cite{IC-AMCNET} with 6 layers is introduced to improve accuracy and a real-time speed is declared, but the model was tested on a GPU and has over 8M parameters, which remain impractical for deployment on embedded and IoT systems that lack GPUs.

Hybrid models, notably CNN-LSTM~\cite{CLDNN} and CNN-GRU~\cite{CGDNet} provide improved robustness and accuracy but at the cost of high computational complexity, making them unsuitable for real-time edge platforms. PET-CGDNN \cite{pet}, while more lightweight ($\sim$ 70k parameters), still has relatively high Multiply-accumulate (MAC) requirement that hinders edge computing, and is also tested on a GPU. Moreover, the reliance on recurrent units limits full exploitation of parallelism.

Overall, existing DL-based AMR networks primarily optimize accuracy, often overlooking deployment constraints. In contrast to edge computing for image tasks, RF signal recognition pipelines have to satisfy sub-millisecond end-to-end latency to keep pace with RF front-end ADC output rates. This makes compact models and hardware acceleration essential. We will discuss related accelerator work in Section~\ref{sec:review_hardware}.

\subsubsection{Covert Channel Detection}
\label{sec:review_cc}
HT-CCs manipulate legitimate RF transmissions to exfiltrate sensitive information \cite{cc_r7,cc_r11,cc_r12,cc_r18,10.1007/978-3-642-36373-3_11,7346830,9933435}. Existing test-based and run-time defenses rely on protocol monitoring, anomaly detection, or handcrafted spectral features. These approaches are typically attack-specific and may fail against sophisticated HT-CCs that mimic normal signal behavior \cite{9933435}.

DL–based detectors have recently been proposed using CNNs on raw I/Q inputs~\cite{alan}, demonstrating improved robustness. However, prior work focuses exclusively on algorithmic accuracy, with limited attention to model compactness, latency, or feasibility for embedded deployment. Hardware-efficient architectures and accelerator support for HT-CC detection remain largely unexplored~\cite{AD-RAS_AIHWHTCC_arXiv26}.

\subsubsection{GNSS Jamming Detection}
\label{sec:review_gnss}

Early GNSS jamming detection relies on handcrafted statistical features and classical machine learning classifiers~\cite{gnss_r22,gnss_r8}, or on covariance/subspace-based mitigation techniques~\cite{gnss_r20}. These methods typically require long observation windows and are computationally intensive. More recent work applies CNNs to spectrogram-transformed representations~\cite{gnss2d} or multimodal CNN–MLP models to raw I/Q samples~\cite{gnssuav} of the same dataset, achieving 90–99\% classification accuracy for multiple jammer types. However, these models often exhibit high inference latency (e.g., several seconds) or rely on GPU evaluation, limiting applicability to real-time embedded systems. 

A number of NN approaches have also been developed for GPS spoofing detection~\cite{gnss_r39,gnss_r55}. Unmanned aerial vehicles (UAVs) platforms are particularly vulnerable to GNSS jamming and spoofing attacks~\cite{gnss_review,gnssuav}, while also facing strict constraints on communication bandwidth, energy, and privacy. As highlighted by edge-computing studies, local inference is essential for timely and autonomous decision-making on drones~\cite{ednet}. Moreover, AMR is also an essential task for UAV communications~\cite{uavamr}. Nevertheless, GNSS jamming classifiers primarily emphasize accuracy and lack designs optimized for real-time, low-power execution on embedded platforms.

\begin{table*}[t]
\centering
\caption{Related work of representative FPGA-based RF AMR hardware accelerator implementations.}
\label{tab:fpga_related}
\setlength{\tabcolsep}{8pt}
\renewcommand{\arraystretch}{1.1}
\begin{tabular}{l c c c c c c c c c}
\hline
\textbf{Work} & 
\textbf{FPGA} & 
\textbf{Dataset} & 
\textbf{NN} & 
\textbf{Weights} & 
\textbf{LUT} & 
\textbf{FF} & 
\textbf{DSP} & 
\textbf{Clock} &
\textbf{Power} \\ 
 & & & \textbf{Type} & \textbf{(\#Bits)} & & & & \textbf{(MHz)} & \textbf{(mW)}\\
\hline
MILCOM'19\cite{Soltani} & XCZU9EG & Private & MLP & 16 & 158,435 & 16,222 & 210 & - & 1,152\\
IPDPSW'20\cite{IPDPSW} & ZCU111 & RadioML2018 & CNN & {2} & {211,000} & {324,000} & {1,407} & 250 & - \\
MILCOM'21\cite{MILCOM} & ZCU104 & RadioML2018 & CNN & {6} & {106,000} & {61,000} & {137} & 250 & - \\
ISCAS'21\cite{emad} & ZCU104 & RadioML2016 & CNN & 16 & 74,680 & 57,726 & 1,116 & 70 & 847\\
IMS'21\cite{tcnn} & ZCU111 & RadioML2018 & CNN & {2} & {61,200} & {92,300} & {391} & 250 & 8,700 \\
IMS'22\cite{jung} & ZCU102 & RadioML2016 & CNN & 16 & 97,900 & 139,200 & 578 & 250 & 10,500\\
MICRO'22\cite{finn} & XCZU28DR & RadioML2018 & CNN & {4} & {229,000} & {131,000} & {0} & 250 & - \\
ICCC'23\cite{kun} & XCZU5EG & RadioML2016 & CNN & 8 & 67,779 & -- & 131 & 200 & 858\\
TAI'24\cite{snn} & PYNQ & RadioML2018 & SNN & 16 & 31,735 & 50,934 & 0 & 137 & 2,167\\
TWC'25\cite{MRTransformer} & XCZU3EG & RadioML2016\&2018 & Transformer & {16} & {43,380} & {29,457} & {328} & -- & 600 \\
TCASAI'25\cite{tcasai} & VIRTEX709 & RadioML2016 & SNN & {16} & {83,572} & {45,060} & {297} & -- & {361}\\
ISCAS'25\cite{song} & ZCU104 & RadioML2018 & CNN & 8 & 75,365 & 88,623 & 1,728 & 115 & 1,191\\
\hline
\end{tabular}
\end{table*}

\subsection{Hardware Accelerators for RF Signal Classification}
\label{sec:review_hardware}

Given the high computational cost of DL models for RF signal recognition (Section~\ref{sec:review_amr}), several hardware accelerators have been proposed to enable low-latency AMR at the edge. Representative designs are summarized in Table~\ref{tab:fpga_related}.

Early work explored multilayer perceptrons (MLPs) for modulation recognition~\cite{Soltani}, achieving low PL (programmable logic) power and latency but supporting only six modulation types and using a private dataset, leaving model generalizability unclear. CNN-based designs using industrial standard RadioML2016/2018~\cite{rml16,rml18} datasets improved accuracy but introduced substantial FPGA cost. The design in~\cite{IPDPSW} achieved ultra-fast throughput but consumed 211k LUTs, 324k FFs, and 1407 DSPs, requiring high-end FPGA boards. Custom RTL pipelines~\cite{MILCOM} improved throughput but still relied on large devices and provided only moderate recognition accuracy.

Low-power accelerators have also been explored. Works such as ~\cite{emad,kun} used 8-16 bit quantization and achieved sub-W power, but were evaluated on the older RadioML2016 dataset with short 128-sample inputs, which do not reflect modern SDR workloads, in which the proposed DL model's parameters would scale significantly with longer input sequences. Conversely, high-accuracy accelerators like~\cite{jung,tcnn} require premium-grade FPGAs and draw $\sim$10W, exceeding many edge-device budgets.

Framework-based accelerators have emerged as well. FINN, a new HLS library which can convert models from deep learning frameworks such as PyTorch to FPGA bitstream, can be used to achieve high throughput AMR~\cite{finn}. However, the designs incur very large resource footprints and target only Xilinx FPGAs.

Neuromorphic computing is a new energy-efficient paradigm, and Spiking Neural Network (SNN)-based AMR~\cite{snn} demonstrated end-to-end low latency AMR but only works with FINN and specific Xilinx FPGA boards such as PYNQ. Another SNN implementation illustrated the energy efficiency of SNNs~\cite{tcasai}.

Transformer-based AMR~\cite{MRTransformer} showed promising accuracy and reduced DSP usage, yet the Vitis AI HLS flow restricts deployment to specific high-end FPGAs, and the reported 2 ms inference latency is insufficient for SDR real-time constraints.
An attention-enhanced CNN~\cite{song} introduced a parallelized on-the-fly streaming architecture with good accuracy–power efficiency tradeoffs, which can be implemented on any FPGA board or ASIC design but still required relatively high resources.

Existing hardware accelerators primarily target a single task (e.g., AMR) and often require large FPGA resources or exhibit limited accuracy under practical Signal-to-Noise Ratio (SNR). They lack mechanisms for task-level adaptation without hardware redesign and do not address RF security tasks. These limitations highlight the need for a flexible, resource-efficient accelerator capable of supporting multiple RF intelligence tasks, leveraging heterogeneous accelerators to reduce resource footprint while maintaining real-time SDR throughput, which is an area that remains largely unexplored.

\section{Proposed Hardware-Efficient AI Model for RF Signal Recognition}

\begin{figure*}[htbp]
\centerline{\includegraphics[width=\textwidth]{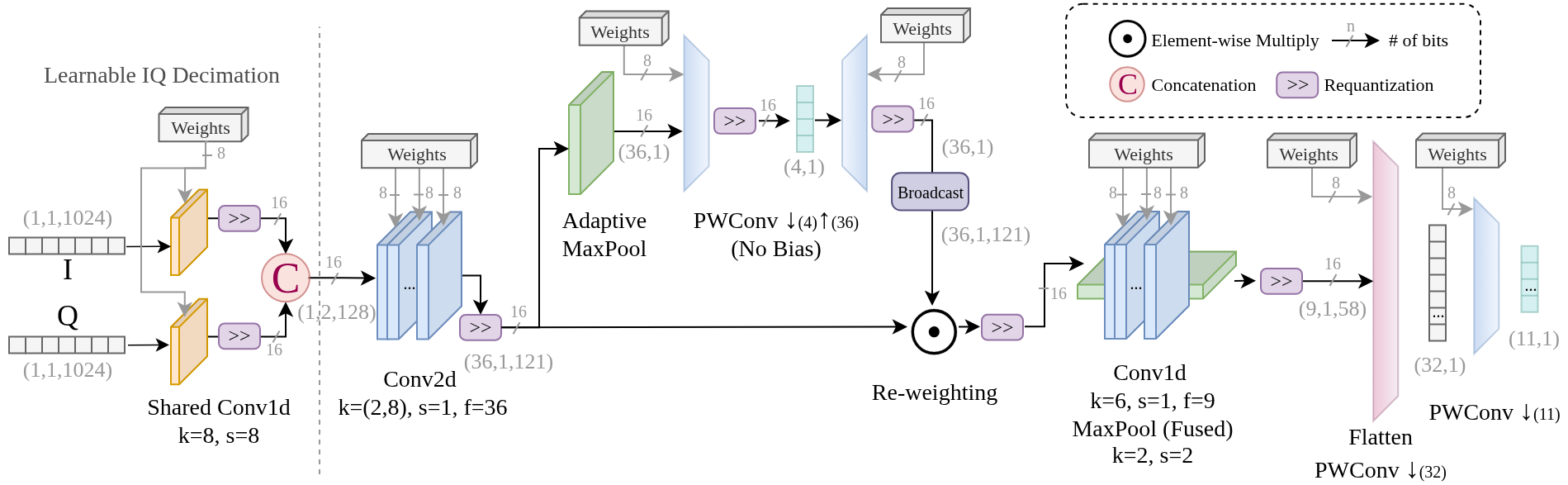}}\vspace{-0.4cm}
\caption{Architecture of the proposed integer-quantized attention-infused CNN model for three RF tasks, namely AMR, HT-CC detection, and GNSS jamming classification. The kernel size, stride, and number of filters are denoted by $k$, $s$, and $f$, respectively. The annotated tensor dimensions are representative of the AMR task using the RadioML2018 dataset. The number of bits of the weights (INT8) and the activations (INT16), is depicted at each layer boundary. The arrows $\downarrow$ and $\uparrow$ denote channel dimensionality reduction and expansion, respectively. Fully connected layers are implemented as pointwise convolutions (PWConv) to map dense layer operations directly onto the hardware convolution engine.}
\label{fig:arch_acnn}\vspace{-0.4cm}
\end{figure*}

Our goal is to identify a single hardware-friendly model that performs effectively across diverse RF signal recognition tasks. The same model architecture is used across all three RF tasks, but for each task, it is trained separately end-to-end, resulting in different weights and hyperparameter tuning. Switching tasks at deployment simply requires loading the corresponding trained weights and reprogramming the hyperparameters at the software level without hardware redesign. Fig.~\ref{fig:arch_acnn} details the data flow in (channel, height, width) format, annotating kernel sizes, strides, and per-layer bit precisions, following PyTorch conventions to ensure straightforward reproduction using DL frameworks.
\vspace{-0.4cm}


\subsection{Learnable Downsampling via the LSDec Module}
\label{sec:LSDec}

Decimation is a standard signal processing technique for reducing sequence length while preserving temporal structure. From a deep learning perspective, we introduce a downsampling operator, implemented as a shared Conv1d kernel applied independently to the I and Q streams. Using a single shared kernel for both channels instead of immediately applying a 2D convolution enforces an inductive bias that encourages symmetric I/Q processing and avoids premature cross-channel mixing~\cite{song}. We set the kernel size $k$ and stride $s$ strictly equal ($k = s = D$), yielding a decimation factor $D$. This reduces the I/Q sequence of length $L$ by a factor of $D$ (yielding a shortened output length of $L/D$), thereby proportionally lowering the computational cost and memory footprint for all subsequent layers. We refer to this component globally as LSDec (Section~\ref{sec:lsdec}).

\subsection{Lightweight Attention-CNN Backbone}
\label{sec:acnnmodel}
Following decimation, the features are processed by a compact Attention-CNN model. Here we summarize the proposed model and its algorithmic adaptations:

\subsubsection{1D-to-2D Spatial Receptive Field}
The decimated I/Q streams are concatenated into a tensor of shape $(1, 2, L/D)$, where the height dimension encodes the I/Q pair. The subsequent Conv2d layer with 36 parallel filters of $2\times8$ kernel size spans both rows, enabling joint I/Q feature extraction. In contrast, applying separate 1D convolutions would isolate the channels and hinder cross-channel feature learning, reducing accuracy. This 2D formulation leverages spatial correlations, analogous to images-based CNNs, and aligns with established RF baselines such as~\cite{emad}.

\subsubsection{Tunable Hardware-Friendly Attention}
We integrate a lightweight channel-attention block between the convolutional layers for feature map calibration. Our tunable SE-style~\cite{senet} attention uses adaptive max-pooling instead of average pooling, and ReLU instead of Sigmoid activation, to ensure hardware-friendly deployment without accuracy loss. The reduction ratio (defaulting from 36 to 4 neurons) is task-configurable, enabling algorithmic adaptation to different RF tasks (e.g., from AMR to HT-CC) for higher accuracy.

\subsubsection{Reduced Filter Count and Pooling}

Compared to the baseline model~\cite{emad,alan} for AMR the first convolution layer (CL) has a reduced number of filters from 45 to 36 and we inserted an additional max-pooling layer after the second convolution, halving the intermediate feature vector length as most parameters originated from the first fully connected layer. These changes reduce parameter count and memory footprint with minimal impact on accuracy.

\subsubsection{Full-Integer Quantization Scheme}
We use PyTorch to train the model end-to-end. Post-training integer quantization is applied to the full model. We propose a specific W8A16-INT format. All weights are quantized to INT8 to minimize the memory footprint and enable efficient hardware multiply-accumulate (MAC) units. Crucially, intermediate activations are maintained at INT16 precision rather than INT8. Because raw I/Q samples exhibit a much higher dynamic range than bounded image pixels, aggressive INT8 activation quantization causes significant classification accuracy degradation. The requantization between layers is achieved purely by right-shift-based scaling. This W8A16-INT format entirely eliminates floating-point units and LUT-based non-linear activation functions from the FPGA datapath, significantly reducing latency, power, and resources.
 

\begin{figure*}[htbp]
\centerline{\includegraphics[width=\textwidth]{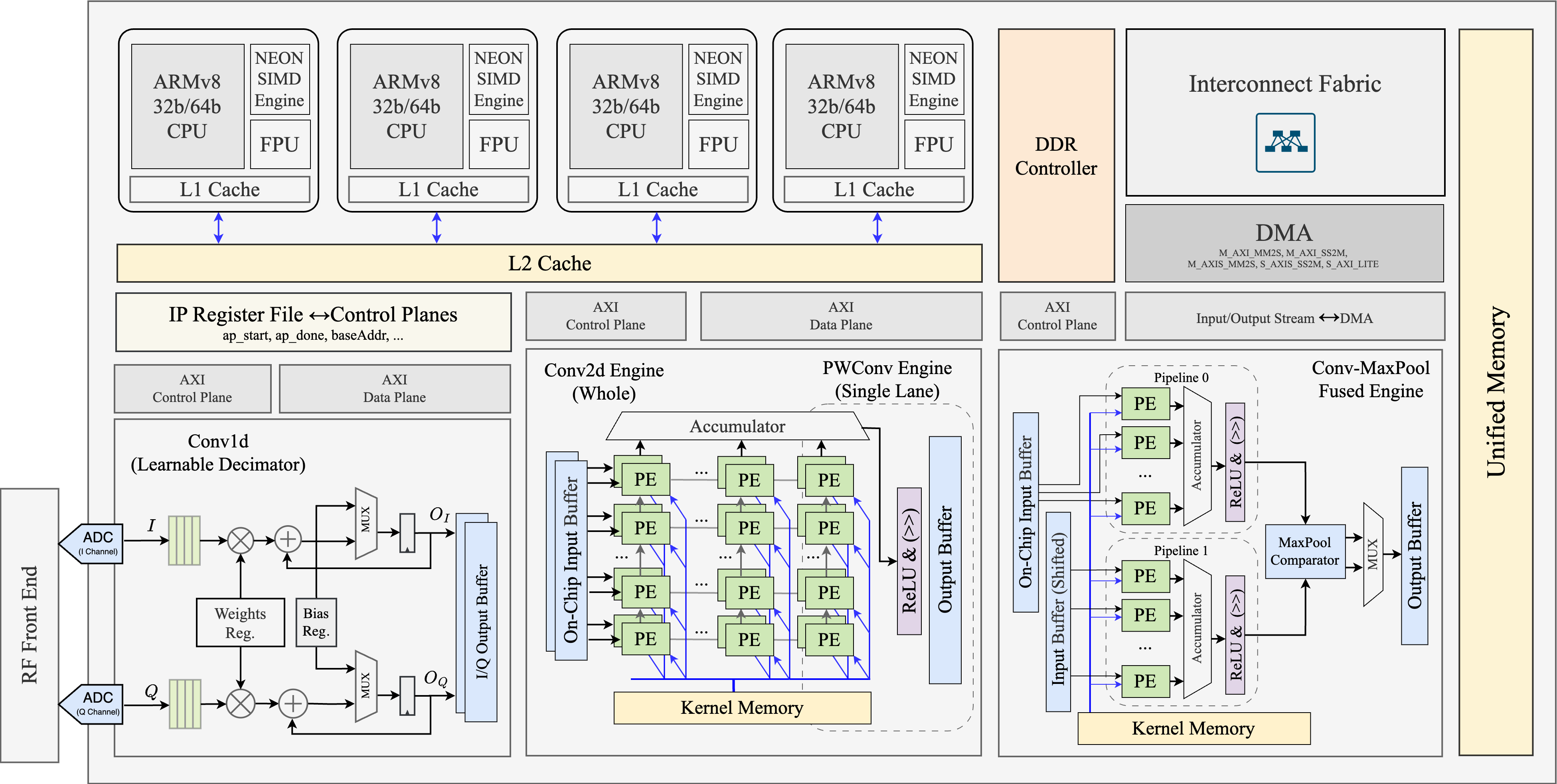}}
\caption{Microarchitecture of the proposed heterogeneous AI accelerator SoC for multi-task RF signal recognition. It integrates a quad-core ARM Cortex-A53 CPU with SIMD engines; the proposed LSDec for task-configurable decimation synchronized with RF front end (clock domain crossing with asynchronous AXI4-Stream FIFO is supported for the subsequent heterogeneous stages); a parallelized Conv2d engine with shared PWConv lanes; a fused Conv-MaxPool engine with dual pipeline implementation. Each accelerator block has its own kernel memory and AXI interface.}
\label{fig:soc}\vspace{-0.2cm}
\end{figure*}

\section{Heterogeneous AI Hardware Accelerator}
\label{sec:hardware_arch}
\subsection{Design Goals and Real-Time Constraints}
Most existing AMR classifiers typically operate with shifting-kernel convolution, requiring full-frame buffering before inference. For high-rate SDR front ends (e.g., AD9361 operating up to 61.44 MSPS), such buffering introduces unavoidable latency and leads to further frame drops.

Our prior streaming accelerator on FPGA~\cite{song} eliminated buffering by using stationary kernels and streaming processing, but required substantial LUT/DSP resources available only on premium FPGAs. The present work generalizes that concept into a heterogeneous SoC suitable for low-power and resource-constrained devices (Fig.~\ref{fig:soc}). The SoC executes several layers cooperatively across an ARM Cortex-A53 CPU cluster (typical IoT and embedded CPU) and a compact pipelined accelerator on FPGA, managed by a lightweight co-execution controller.

To capture wideband spectral features, the RF front-end operates at a sample rate of $F_s$, streaming raw I/Q samples directly to the AI SoC as previously depicted in the system overview in Fig.~\ref{fig:system}. In heterogeneous embedded platforms, the CPU typically exhibits availability due to the event-driven nature of sensing workloads~\cite{idle}. To leverage this, we partition the workload between the FPGA and CPU. The design goal is to balance this load to achieve a sub-100 $\mu$s inference latency (approx. $>$10 kFPS), ensuring high-density spectrum monitoring while utilizing cooperative heterogeneous resources.

\subsection{Learnable Streaming Decimator (LSDec)}
\label{sec:lsdec}
\begin{figure}[t]
\centerline{\includegraphics[width=1\linewidth]{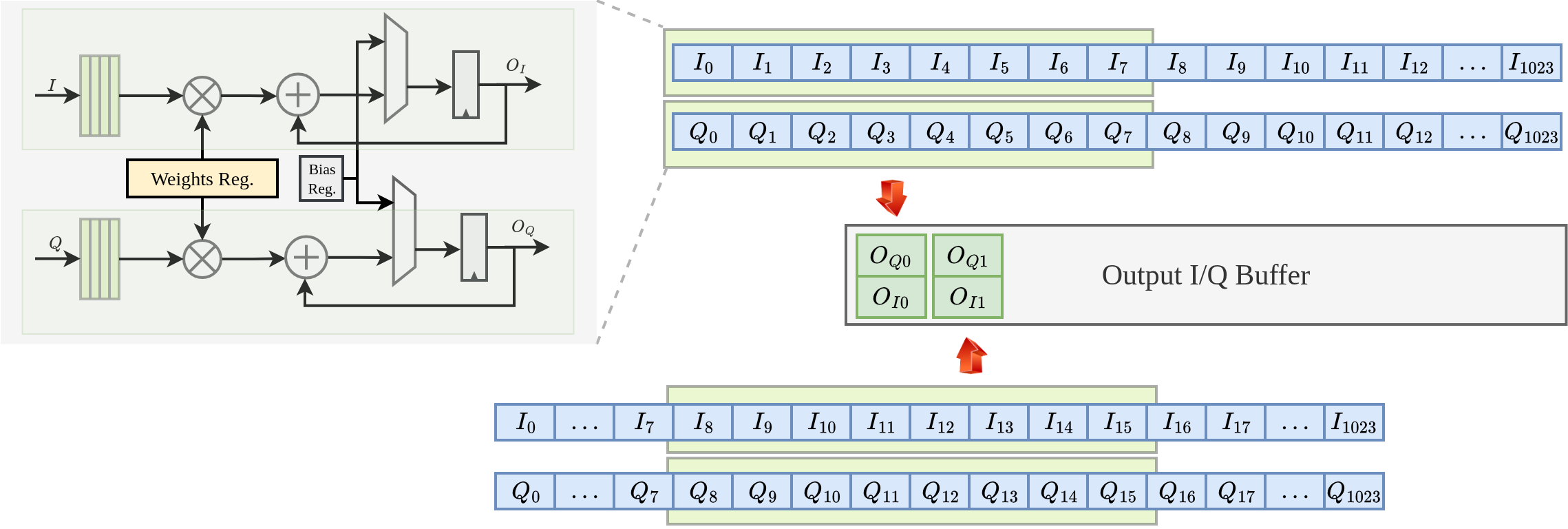}}
\caption{Hardware architecture of LSDec. Utilizing a shared, stationary weights kernel for the I and Q channels, the module exploits the non-overlapping property of stride $s=k=D$. This allows a single MAC pair per channel to process incoming ADC samples every clock cycle. A valid decimated output is generated, and the accumulator is reset, every $D$ clock cycles.}
\label{fig:decimator}\vspace{-0.4cm} 
\end{figure}

The learnable decimator is implemented as a stationary Conv1d kernel interfacing directly with the high-speed RF ADC. A single shared set of $k=D$ weights is applied independently to the I and Q channels, performing decimation and feature extraction on the fly. As shown in Fig.~\ref{fig:decimator}, unlike shifting-kernel convolutions, the stationary design exploits the non-overlapping property of stride $s=k=D$. Consequently, as each ADC sample arrives per clock cycle, it is multiplied by its corresponding weight and added to a running accumulator. After $D$ clock cycles, the accumulator yields one valid output and resets. This eliminates shift registers, requiring only one MAC unit per channel and bypassing full-frame buffering.

\subsection{Accelerator Microarchitecture}
Fig.~\ref{fig:soc} shows the proposed heterogeneous accelerator with unified memory architecture. To maximize data locality during CPU execution, feature maps and CPU-side weights are stored in DRAM in 1D row-major order (RMO), aligning naturally with the Cortex-A53 cache hierarchy and SIMD instructions. In contrast, the accelerator operates on 3D ($C \times H \times W$) tensorized feature maps. The conversion is explained in Section~\ref{sec:memarch}.

\subsubsection{Conv2d Engine}

Pretrained INT8 weights are partitioned in on-chip memory. The first Conv2d layer partitions its $N=36$ filters such that $K_a$ filters execute on the accelerator and $N-K_a$ on the CPU. The hardware datapath is fully pipelined (input registering, multiplier stage, accumulator). After the initial pipeline fill, each processing element (PE) sustains a continuous throughput of one valid MAC operation per clock cycle. The CPU and accelerator operate in parallel on mutually exclusive filter subsets, writing to the shared memory. The dynamic workload distribution across available CPU cores will be discussed in Section~\ref{sec:algo}.

\subsubsection{PWConv Engine (Folded)}
PWConv is mathematically equivalent to dense layers; its operations can be folded into the convolution engine by interpreting neurons as channels. Reusing the existing Conv2d pipelined datapath completely eliminates the need for dedicated dense-layer IPs, strictly reducing area and power, with only a lightweight control signal that configures the engine between the inference phases. The overhead is negligible compared to a separate dense layer IP. A quantitative discussion will be presented in Section~\ref{sec:hw_results}. Lightweight, task-configurable Attention layers are executed entirely on the CPU to enable adaptation (e.g., modifying the reduction ratio) without accelerator IPs.

\subsubsection{Conv-MaxPool Fused Engine}
The second convolution and subsequent max-pooling stages dominate computational cost. Executing them separately and sequentially using two separate accelerator IPs would require writing and re-reading intermediate feature maps, incurring significant latency and DDR bandwidth penalties with extra roundtrip and additional AXI interfaces. As illustrated in Fig.~\ref{fig:soc}, our fused dual-pipeline engine fetches a sliding window using shifted on-chip buffers to process two adjacent spatial outputs concurrently. A hardware comparator immediately evaluates the outputs on the fly, storing only the maximum value to streamline the pooling operator. This halves the memory traffic and eliminates intermediate transactions.

\subsubsection{Data Path and Memory Architecture}
\label{sec:memarch}

The heterogeneous SoC connects the streaming RF domain and the shared-memory execution domain (Fig.~\ref{fig:system}). To bridge the high-speed RF ADC clock domain, LSDec utilizes a dual-path broadcast. A point-to-point AXI4-Stream interface feeds the decimated samples directly to the Conv2d Engine, bypassing DDR roundtrip to minimize pipeline latency. Concurrently, an AXI master interface with
an asynchronous FIFO safely burst-writes the decimated sequence into the unified DDR memory as 1D row-major array.

Data layout transformations (tensorization) are handled seamlessly at the hardware IP boundaries with zero software overhead. While the CPU operates on 1D flat arrays with row-major indexing with direct SIMD acceleration, the receiving logic inside the accelerator's on-chip buffers utilizes hardware line buffers and shift registers to demultiplex the 1D AXI stream into the 3D parallel registers required by the PEs. The kernel memories of the accelerator engines are implemented as fully partitioned, distributed on-chip memory located physically adjacent to the PEs, enabling single-cycle parallel weight access during inference. To achieve shared-memory execution for distributed layers, the CPU and accelerator operate in parallel on mutually exclusive convolutional filter subsets (or neurons, in the case of PWConv/Dense layers). The hardware accelerator bypasses the CPU cache, utilizing the DMA engine to read inputs and write outputs directly to the physical DDR. Concurrently, the CPU fetches the exact same inputs from the unified memory through its native L1/L2 cache hierarchy, distributing the workload across available cores based on availability, and writes its assigned output channels back to the DDR via the cache.
  
\subsection{Co-scheduling Algorithm and Cache Coherency}

Let a convolution/dense layer with $N$ output channels produce output vector: $\textbf{y} = \{y_0,y_1,...,y_{N-1}\}$, where each $y_i$ corresponds to either an output feature map of a convolution layer or a neuron output of a dense layer. Let $t_a$ and $t_c$ denote per-channel execution latency on accelerator and CPU, respectively.

As the CPU and accelerator initiate inference in parallel, total latency is: 
$$T(K_a) = \max (K_at_a, (N-K_a)t_c)$$

The unconstrained minimizer is computed when the two finish simultaneously:
$$K_at_a = (N-K_a) t_c \rightarrow K_a^* = \frac{Nt_c}{t_a+t_c} $$

However, shared-memory execution introduces cache-line boundary hazards: if CPU and accelerator write to addresses inside the same cache line, stale data or premature evictions can occur. The ARM-based CPU has 64-byte cache line size and if the CPU and accelerator update data that fall inside the same cache line, the CPU cache will reload data unexpectedly or evict accelerator-written entries, causing corruption and nondeterminism.

Let each channel output $y_i$ occupy a contiguous memory segment of $B$ bytes and let the CPU cache line size be $L_c$. The boundary address $\Delta = K_a B$ must align to a cache line, e.g., ($K_aB$) mod $L_c$ = 0.

Thus the feasible set is defined as:
$$\mathcal{K} = \{K_a \in {0,...N}: (K_a B) \ \text{mod} \ L_c =0 \} $$

We seek the aligned partition that minimizes the total execution time
$$K^*_a = \text{argmin}_{K_a \in \mathcal{K}} \ T(K_a)$$

As the latency equation is convex in $K_a$, and $\mathcal{K}$ contains discrete candidates defined by ($K_a B$) mod $L_c$ = 0, an enumeration of $\mathcal{K}$ yields the optimal solution. 

 The proposed CPU-accelerator co-scheduling algorithm is illustrated in Alg.~\ref{alg:cosched}. It is designed to be portable across different embedded CPU architectures. On our ARM Cortex-A53 processor, a particular subset of layers is offloaded to the CPU; however, when deployed on CPUs with different performance characteristics, the algorithm automatically adjusts the number of layers to be offloaded. A case study of this execution will be presented in Section~\ref{sec:hw_results}. 

\label{sec:algo}
\begin{algorithm}[t]
\caption{Cache-Aligned Heterogeneous Co-Scheduling}
\label{alg:cosched}
\begin{algorithmic}[1]

\STATE \textbf{Input:} $N$, $B$, $L_c$
\STATE \textbf{Output:} $K_a$

\STATE $t_a \gets \text{ProfileAcceleratorLatency}()$
\STATE $t_c \gets \text{ProfileCPULatency}()$

\STATE $K_{\text{ideal}} \gets \frac{N t_c}{t_a + t_c}$

\STATE $\mathcal{K} \gets \emptyset$

\FOR{$K = 0$ to $N$}
    \IF{$(K B) \bmod L_c = 0$}
        \STATE $\mathcal{K} \gets \mathcal{K} \cup \{K\}$
    \ENDIF
\ENDFOR

\STATE $T_{\min} \gets \infty$, \ $K_a \gets 0$

\FOR{$K \in \mathcal{K}$}
    \STATE $T \gets \max(K t_a,\ (N-K)t_c)$
    \IF{$T < T_{\min}$}
        \STATE $T_{\min} \gets T$
        \STATE $K_a \gets K$
    \ENDIF
\ENDFOR
\STATE \RETURN $K_a$

\end{algorithmic}
\end{algorithm}

Our framework integrates Alg.~\ref{alg:cosched} with runtime adaptive profiling (RAP) to manage CPU availability. The latencies $t_a$ and $t_c$ can be profiled statically at deployment or updated dynamically with each frame via the ARM CPU cycle counter register, readable in a single instruction with negligible overhead. The algorithm natively extends to multi-core topologies. If $M$ symmetric CPU cores are allocated to the RF tasks, the expected aggregate per-channel latency is estimated as $t_c/M$, since filter computations are independent across output channels. This linear scaling model provides an accurate initial partition; measured latency from subsequent frames refines the estimate further. The feasible set $\mathcal{K}$ is further constrained by a core-symmetry condition: $(N - K) \bmod M = 0$, ensuring an evenly distributed software workload across cores. Where cache-line alignment restricts the feasible set, memory padding can be applied to force the output channel byte size to a multiple of $L_c$ to unlock unconstrained fine-grained partitioning. The framework operates entirely at the application layer, and partition updates take effect at I/Q frame boundaries: updating $K_a$ requires only writing new channel-start index values to each core's parameter block and a single DMA transfer descriptor write, with overhead negligible relative to the 98 $\mu s$ frame period. In this work, to guarantee strict real-time deterministic execution, we evaluate a worst-case scenario where a single core is allocated for RF tasks (Section~\ref{sec:hw_results}). In deployments, the framework automatically offloads more filters to CPU, reducing the accelerator's dynamic power proportionally to the fraction of idle PEs.

Regarding synchronization and data path coherency, the co-execution controller operates a lightweight state machine: (1) waits for both the CPU and accelerator completion flags; (2) issues a targeted data cache clean-and-invalidate instruction over the CPU-written memory region; (3) waits for all dirty cache lines to be physically written to the shared DDR; (4) signals the DMA engine to begin reading the merged feature map for the subsequent layer. Because the co-scheduling algorithm guarantees that the partition boundary $K_aB$ is strictly cache-line aligned, the CPU and accelerator write to mutually exclusive physical cache lines, completely eliminating false-sharing hazards and ensuring the flush covers only the CPU-written region. The synchronization overhead is therefore bounded by the number of CPU-written cache lines; a quantitative analysis of this worst-case overhead is provided in Section~\ref{sec:caseco}.

\section{Experimental Setup}
\subsection{Datasets}

We evaluate the proposed heterogeneous AI accelerator on three representative RF tasks: AMR, HT-CC detection, and GNSS jamming classification, using three publicly available datasets that differ in signal structure, frame length, and SNR conditions, enabling a comprehensive assessment of generalizability.

\subsubsection{AMR RadioML2018} We use the RadioML2018 dataset~\cite{rml18} for the AMR task. It provides 2{,}555{,}904 frames of over-the-air RF captures spanning SNRs from -20 to +30 dB. We follow the 50\%--25\%--25\% training/validation/test split used in~\cite{dae}, and evaluate on the 11 normal modulation classes (FM, GMSK, OQPSK, BPSK, 8PSK, AM-SSB-SC, 4ASK, AM-DSB-SC, QPSK, OOK, and 16QAM) as in ~\cite{dae,snn}. Each frame consists of $2\times 1024$ I/Q samples.

\subsubsection{HOST HT-CC}
The HT-CC dataset~\cite{alan} contains over-the-air IEEE 802.11 transmissions collected via a bladeRF SDR, comprising four representative HT-CC mechanisms spanning PHY-layer, synchronization, dirty-constellation, and analog front-end attacks, along with a clean reference class. Each SNR point (1–20 dB in 4 dB steps) includes 2,000 frames of size $2\times640$, totaling 80k samples. Due to the smaller dataset size, we use an 80\%–-10\%–-10\% split for training/validation/testing.

\subsubsection{GNSS Jamming}
We use a publicly available GNSS-band interference dataset~\cite{jammingdataset} containing six classes: no-interference and five jammer types (continuous-wave, FM, chirp, narrowband, and pulsed). Frames consist of raw complex baseband samples generated across controlled SNR conditions. The dataset includes 6,000 training and 1,500 test samples and is used to evaluate cross-domain generalizability due to its distinct spectral structure.

\subsection{AI and Hardware Development Setup}

For fair comparison and reproducibility, all models (this work and baselines), are trained in the same software environment on an NVIDIA A100 GPU using identical training, validation, and test splits. All models are optimized end-to-end using the Adam optimizer with a learning rate of $10^{-3}$, and cross-entropy loss. The proposed hardware was synthesized and implemented on a Zynq™ UltraScale+™ MPSoC ZCU104 Evaluation board. FPGA synthesis uses Vivado 2025.1, and hardware–software co-design is performed in Vitis Unified IDE 2025.1. A summary of the development platform is provided in Table~\ref{tab:setup}.

\begin{table}[t]
    \centering
    \setlength{\tabcolsep}{10pt}
    \caption{Model training and hardware development environment}
    \begin{tabular}{ll}
        \hline
        \textbf{Category} & \textbf{Details} \\
        \hline
        GPU & NVIDIA A100 80 GB PCIe \\
        CPU & Intel\textsuperscript{\textregistered} Xeon\textsuperscript{\textregistered} Gold 6300 @ 2 GHz \\
        RAM & 128 GB \\
        Operating System & Ubuntu 24.04.2 LTS \\
        PyTorch / Python &  2.7.1 / 3.11.14 \\
        CUDA  & 12.8 \\
        \hline 
        FPGA & Zynq™ UltraScale+™ MPSoC ZCU104 \\ 
        Xilinx Vivado \& Vitis  & v2025.1 \\ 
        \hline
    \end{tabular}
    \label{tab:setup}\vspace{-0.3cm}
\end{table}

\section{Results and Discussion}
\label{sec:results}
\subsection{Model Comparison and Analysis}
We conducted a comparative analysis of various models. The selected baselines include both AMR-specific models~\cite{CGDNet,emad,dae,IC-AMCNET,CLDNN,lstm2amr,pet}, and widely adopted general purpose edge-oriented models, such as MobileNet~\cite{mobilenetv2} and MobileViT~\cite{mobilevit}, which are included since recent studies formulate AMR as an image-like learning problem, enabling the reuse of vision-oriented architectures. 

We first train and evaluate the models on the AMR task, with results summarized in Section \ref{susubsec:AMR_task} and Table~\ref{tab:model_comparison_rml}. Because specialized DL models for HT-CC and GNSS jamming remain underexplored (Sections~\ref{sec:review_cc} and \ref{sec:review_gnss}), and all three datasets are comprised of raw I/Q sequences aligned with our end-to-end inference goal, we conduct further experiments by training and evaluating the existing AMR baselines on the two additional RF tasks to comprehensively benchmark the proposed model against the baselines.

The proposed Attention-CNN model serves as a unified backbone across all three RF tasks. At deployment, specific tasks are instantiated simply by loading their corresponding pre-trained weights from the shared DDR. If baseline accuracy on a new task is suboptimal, the model is easily reconfigured in software by adjusting the attention reduction ratio or the number of neurons in the first dense layer. For example, because our default hyperparameters were optimized for AMR, we applied this software-level tuning for the HT-CC dataset (denoted as "Task Tuned" in Table~\ref{tab:model_comparison_htcc}), whereas the GNSS-Jamming task achieved high accuracy without modification. Crucially, these adjustments are handled entirely by the CPU and the Conv2d Engine without requiring any hardware redesign. The three tasks can be scheduled sequentially, as the information is carried by all frames. This  high-rate task-multiplexing is possible thanks to the 98 $\mu$s per-frame inference latency and the small weight count of the model which makes model updates very fast. To the best of our knowledge, no prior RF hardware accelerator has demonstrated deployment flexibility across distinct RF intelligence tasks (spectrum monitoring, security, navigation) on a single hardware instance using a unified AI architecture.

\begin{table}[t]
\setlength{\tabcolsep}{10pt}
\caption{Comparison of AI models on RadioML2018 dataset (SNR -20 to 30 dB). All models are trained on the same GPU using identical training/validation/test splits except~\cite{snn}. Average accuracy across all available SNR values is reported.}
\label{tab:model_comparison_rml}
\centering
\begin{tabular}{l c c c}
\hline
\textbf{Model} & \textbf{Accuracy} & \textbf{\#Param} & \textbf{kMACs} \\
\hline
CGDNet~\cite{CGDNet}       & 71.1\% & 662,533   & 171,541 \\
SNN (IQ)~\cite{snn}         & 64.3\% & 83,000    & --      \\
CNN (IQ)~\cite{emad}       & 70.1\% & 295,055   & 3,648   \\
DAE~\cite{dae}             & 67.3\% & 14,989    & 14,094  \\
CLDNN2~\cite{CLDNN}        & 68.6\% & 696,843   & 504,681 \\
IC-AMCNet~\cite{IC-AMCNET} & 72.9\% & 8,604,043 & 119,638 \\
LSTM (IQ)~\cite{lstm2amr}       & 74.9\% & 201,088   & 207,095 \\
MobileNetv2~\cite{mobilenetv2} & 74.9\% & 2,237,387 & 196,835 \\
MobileViTv2-0.5~\cite{mobilevit}& 75.5\% & 1,116,036   & 142,521 \\
PET-CGDNN~\cite{pet}       & 74.1\%  & 71,614    & 72,123  \\
 \rowcolor{gray!20} 
    This Work   & 73.3\% & 19,961    & 346     \\
\hline
\end{tabular}
\end{table}
\subsubsection{AMR\label{susubsec:AMR_task}}

Existing approaches for the AMR task predominantly prioritize accuracy at the expense of model size and computational complexity, neglecting the fact that RF signal classification problems are latency-critical and require local inference to satisfy speed and data privacy requirements. As shown in Table~\ref{tab:model_comparison_rml}, in terms of average accuracy across all SNRs, CGDNet~\cite{CGDNet} achieves 71.1\% accuracy on the RadioML2018 dataset but requires over 660k parameters and a substantial computational cost. This trend becomes more pronounced in IC-AMCNet \cite{IC-AMCNET}, which reaches 72.9\% accuracy while employing more than 8.6M parameters. Recurrent architectures further exacerbate this issue: the LSTM-based model in \cite{lstm2amr} attains a higher accuracy of 74.9\%, yet incurs significantly higher kMACs and suffers from inherent sequential dependencies, limiting parallelism and throughput in hardware implementations. 

\begin{figure}[t]
\centerline{\includegraphics[width=\linewidth]{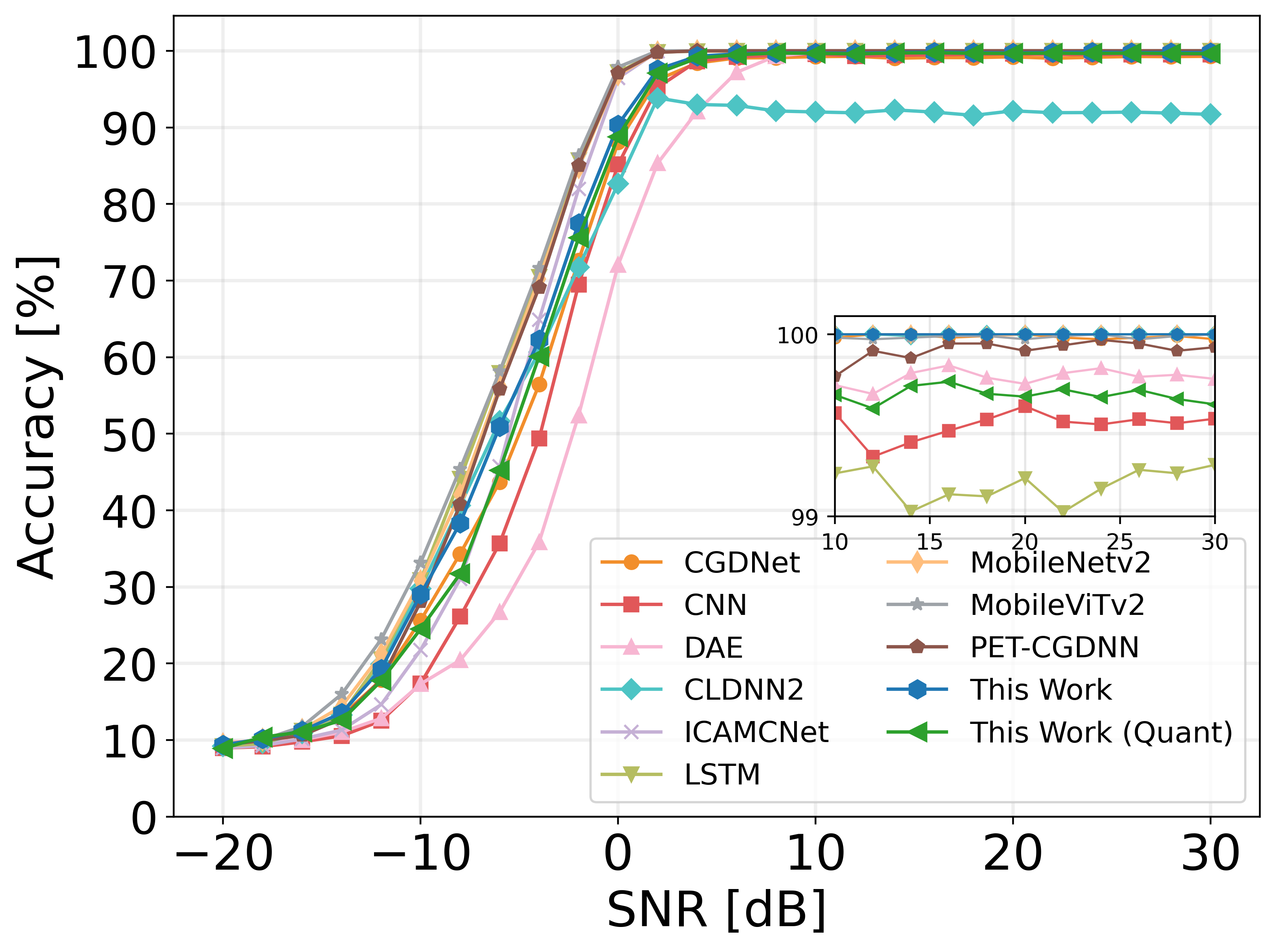}}
\caption{Accuracy vs. SNR curve on the RML2018 dataset for normal classes. Quantization reduces accuracy, but when the SNR is relatively high, the difference is minimal.}
\label{fig:curve_rml2018}
 \centerline{\includegraphics[width=\linewidth]{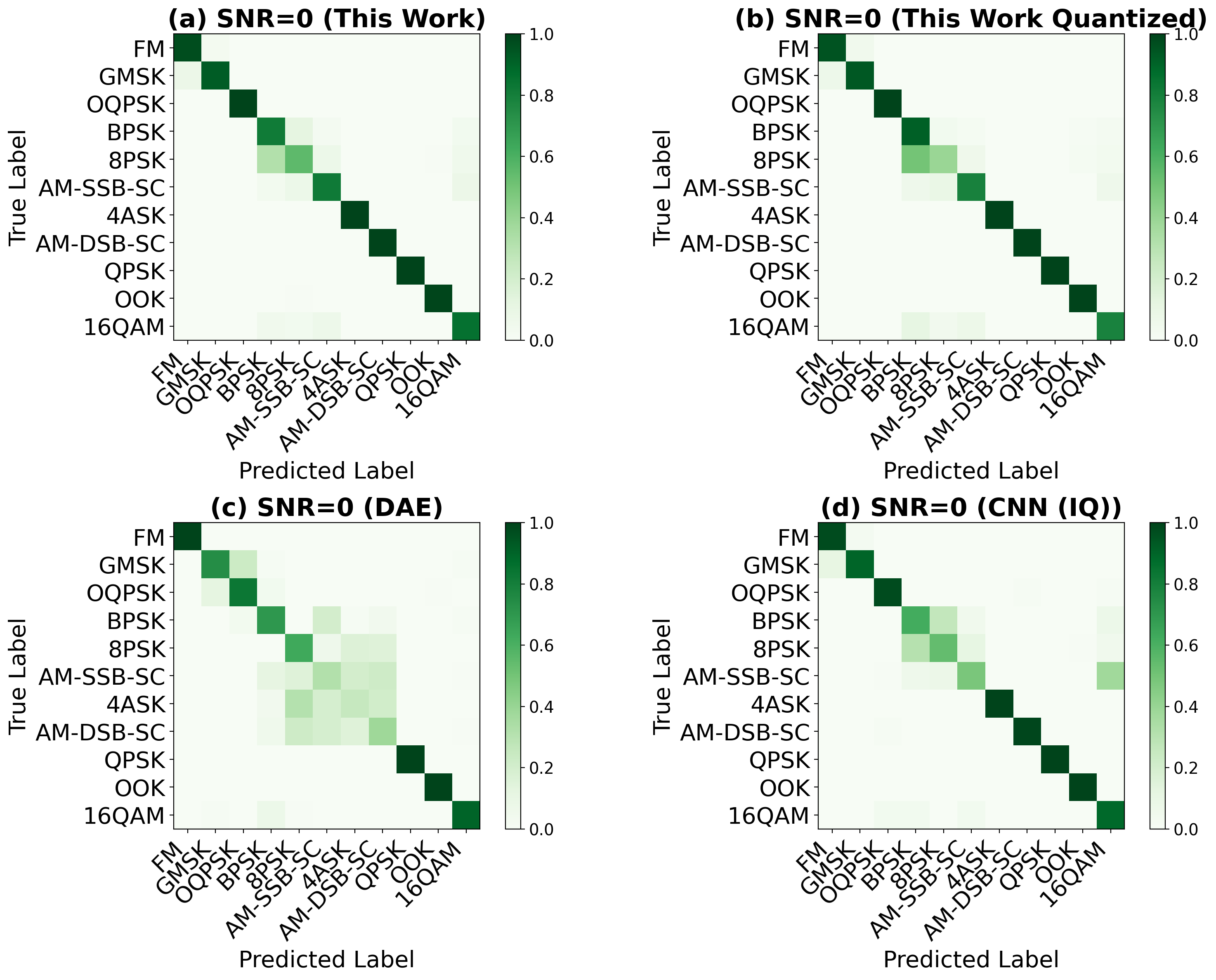}}
\caption{The confusion matrix at SNR = 0 dB for the AMR task using RadioML2018 dataset. Comparison of this work compared to the most lightweight model in the literature (DAE) and the baseline CNN: (a) This work, (b) This work (quantized), (c) DAE~\cite{dae} (d) CNN(IQ)~\cite{emad}.}
\label{fig:confmat}\vspace{-0.2cm}
\end{figure}
\begin{figure}[t]
\centerline{\includegraphics[width=\linewidth]{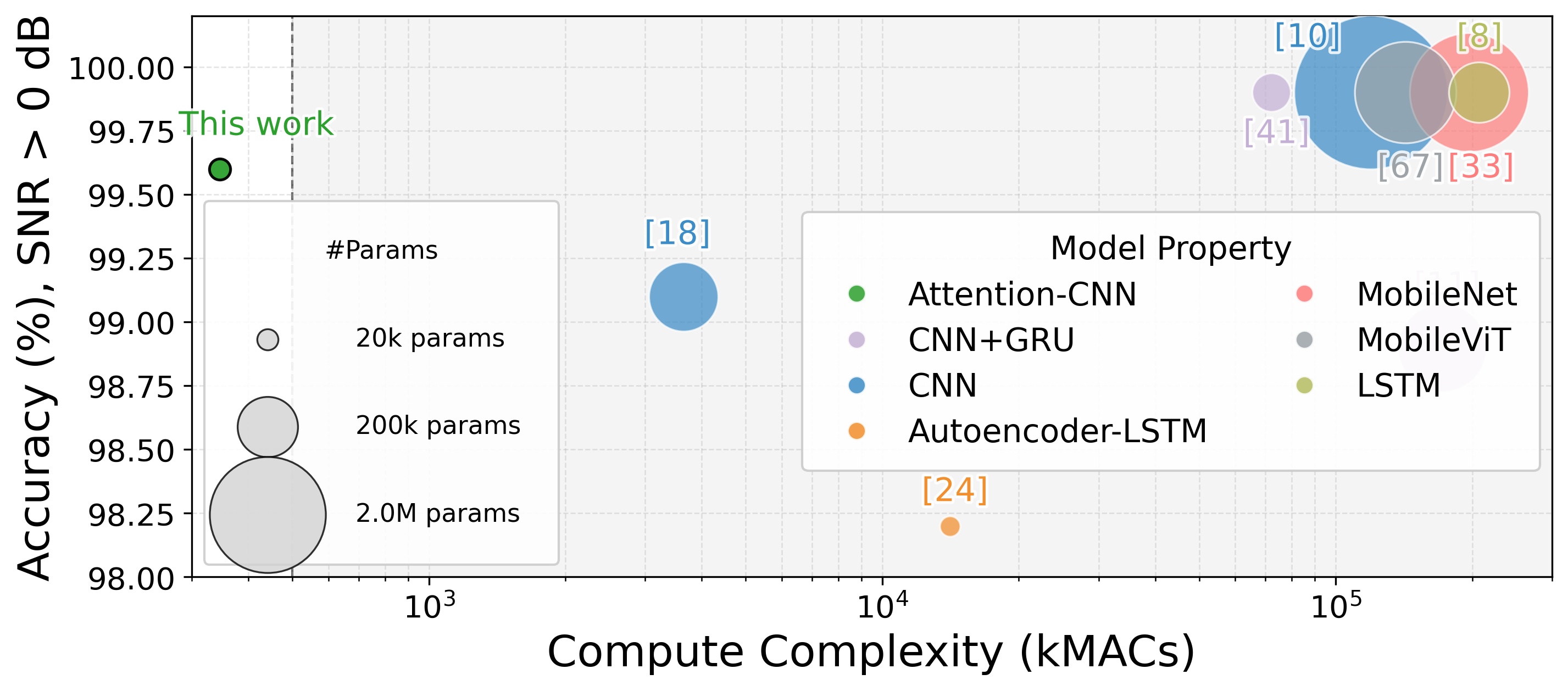}}
\caption{Figure of merit: accuracy–complexity trade-off for AMR measured as average accuracy (SNR $>$ 0 dB) versus kMACs. Log-scale is applied on the complexity axis to visualize both the huge gap between the proposed model and prior work, and the small variations among certain existing models.}
\label{fig:fom}\vspace{-0.4cm}
\end{figure}

General-purpose edge models (MobileNet~\cite{mobilenetv2} and MobileViT~\cite{mobilevit}) demonstrate strong accuracy when trained on the RadioML2018 dataset. However, they require millions of parameters and hundreds of thousands of kMACs. However, these architectures are primarily designed for vision workloads, where end-to-end latencies on the order of several milliseconds are acceptable. In contrast, RF signal recognition operates under strict real-time constraints as the data is streamed from the SDR front-end ADC. These translate into excessive FPGA resource utilization or ASIC area and power consumption, making them impractical for embedded and IoT platforms without premium mobile SoCs.

Several lightweight models attempt to address deployability concerns. DAE~\cite{dae} is an LSTM-Autoencoder that employs fewer than 15k parameters, comparable to the proposed design; however, its reliance on recurrent layers results in lower parallelism and higher computational complexity, and the accuracy is limited at 67.3\%. PET-CGDNN \cite{pet}, a hybrid CNN–GRU architecture, achieves a relatively high accuracy of 74.1\% while reducing parameters and kMACs compared to many deep CNN counterparts, yet its computational cost remains substantially higher than that of the proposed model.

In contrast, the proposed model achieves a balanced tradeoff between accuracy, model compactness, and computational efficiency. It attains 73.3\% average accuracy across all SNRs, outperforming the baseline CNN \cite{emad} by 3.2\% while reducing parameter count by 93.2\%. Compared with the lightweight DAE \cite{dae}, the proposed model improves accuracy by 6\% while requiring only 346 kMACs, orders of magnitude fewer operations than most competing approaches. 


\begin{table}[t]
\setlength{\tabcolsep}{8pt}
\caption{Comparison of AI models on the HT-CC dataset (SNR 1 to 29 dB). All models are trained on the same GPU using identical training/validation/test splits. Average accuracy across all available SNR values is reported.}
\label{tab:model_comparison_htcc}
\centering
\begin{tabular}{l c c c}
\hline
\textbf{Model} & \textbf{Accuracy} & \textbf{\#Param} & \textbf{kMACs} \\
\hline
CGDNet~\cite{CGDNet}       & 76.8\% & 660,991   & 130,298 \\
CNN (IQ)~\cite{emad}       & 91.5\% & 184,265   & 2,265   \\
DAE~\cite{dae}             & 79.5\% & 14,887    & 8,810  \\
CLDNN2~\cite{CLDNN}        & 93.4\% & 619,269   & 314,916 \\
IC-AMCNet~\cite{IC-AMCNET} & 92.0\% & 5,457,541 & 74,774 \\
LSTM (IQ)~\cite{lstm2amr}       & 88.8\% & 200,320   & 129,434 \\
MobileNetv2~\cite{mobilenetv2} & 95.2\% & 2,229,701 & 123,020 \\
MobileViTv2-0.5~\cite{mobilevit}& 95.2\% & 1,114,494   & 89,075 \\
PET-CGDNN~\cite{pet}       & 84.6\% & 70,840    & 44,785  \\
    This Work   & 88.8\% & 19,760    & 345     \\
 \rowcolor{gray!20} 
    This Work (Task Tuned) & 90.1\% & 29,078    & 354     \\
\hline
\end{tabular}\vspace{-0.5cm}
\end{table}

\begin{figure}[t]
\centerline{\includegraphics[width=\linewidth]{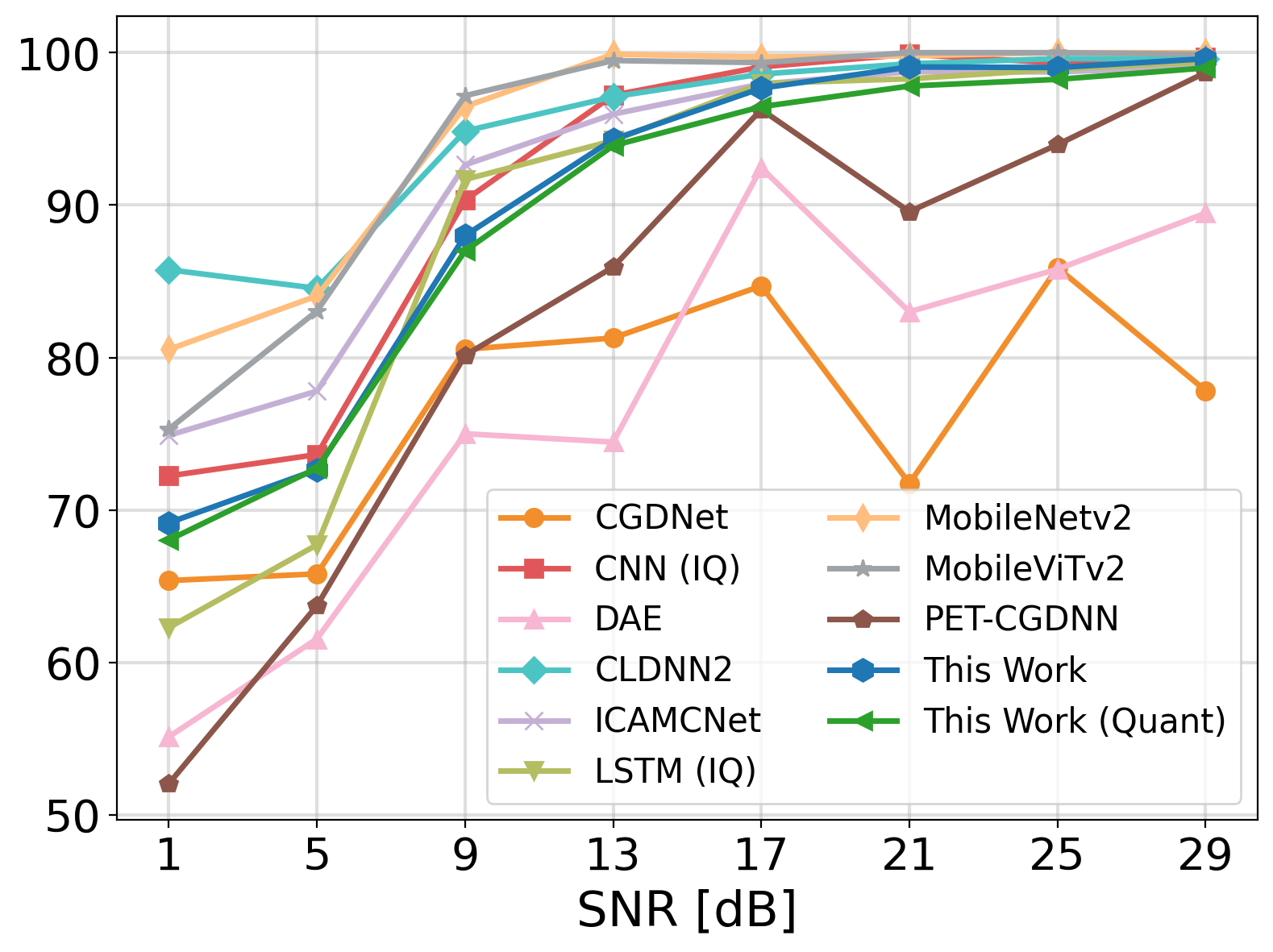}}\vspace{-0.4cm}
\caption{Accuracy vs. SNR curve on the HT-CC dataset.}
\label{fig:curve_htcc}\vspace{-0.2cm}
\end{figure}

\begin{table}[t]
\setlength{\tabcolsep}{8pt}
\caption{Comparison of AI models on GNSS Jamming dataset (one fixed SNR and no sweeping). All models are trained on the same GPU (except~\cite{gnssuav}) using identical training/test splits.}
\label{tab:model_comparison_gps}
\centering
\begin{tabular}{l c c c}
\hline
\textbf{Model} & \textbf{Accuracy} & \textbf{\#Param} & \textbf{kMACs} \\
              
\hline
CGDNet~\cite{CGDNet}       & 92.1\% & 661,248   & 1,819,486 \\
DAE~\cite{dae}             & 99.9\% & 14,904    & 225,258  \\
CNN (IQ)~\cite{emad}       & 99.9\% & 4,714,127 & 58,886   \\
CLDNN2~\cite{CLDNN}        & 97.7\% & 3,764,998  & 8,087,317 \\
IC-AMCNet~\cite{IC-AMCNET} & 99.7\% & 134,301,446 & 1,912,307 \\
MobileNetv2~\cite{mobilenetv2} & 99.9\% & 2,230,982 & 3,147,701\\
MobileViTv2-0.5~\cite{mobilevit}& 99.9\% & 1,114,751   & 2,279,123 \\
PET-CGDNN~\cite{pet}       & 99.8\% & 71,614    & 1,164,478  \\
MLP+CNN~\cite{gnssuav}       & 99.0\% & -    & - \\
 \rowcolor{gray!20} 
    This Work   & 99.5\% & 19,793    & 374     \\
\hline
\end{tabular}\vspace{-0.5cm}
\end{table}

Compared to lightweight models such as DAE \cite{dae}, the proposed model also exhibits greater robustness in low-SNR conditions. As shown in Fig. \ref{fig:curve_rml2018}, it achieves 90\% accuracy starting from 0~dB and exceeds 99\% accuracy above 4~dB, while DAE requires 4~dB to exceed 90\% and 8~dB to reach 99\%, a consistent 4~dB SNR advantage, highlighting the effectiveness of the learnable decimation coupled with attention recalibration in enhancing discriminative feature extraction under higher noise. This improvement is supported by three pieces of evidence. First, the SNR curves in Fig.~\ref{fig:curve_rml2018} confirm this advantage across lower SNR regime compared to lightweight baselines. Second, the confusion matrix at SNR~=~0~dB (Fig. \ref{fig:confmat}) shows that the proposed model better separates spectrally similar modulation pairs that lightweight baseline models confuse under noise, reflecting the attention mechanism's ability to recalibrate channel features toward higher discriminative signal characteristics, and showing improved classification fidelity at lower SNR by comparing with the most lightweight model in the literature, e.g. DAE ~\cite{dae}, and a baseline CNN with the same number of convolutional and dense layers~\cite{emad,alan}. Third, the ablation study which we will detail later in Section~\ref{subsec:ablation_study}, isolates each component's contribution: LSDec alone improves accuracy from 70.1\% to 71.9\%, and adding attention further raises it to 73.3\%, confirming both components contribute independently to improved recognition under challenging SNR conditions.

A figure of merit comparing AMR average accuracy (for SNR $>$ 0 dB) against computational complexity is shown in Fig.~\ref{fig:fom}. The proposed model is markedly more efficient, requiring only 346 kMACs for the AMR task and remaining below 500 kMACs across all three RF tasks, while still achieving high recognition accuracy. This highlights the favorable accuracy–efficiency balance of the proposed architecture relative to existing approaches.

\subsubsection{HT-CC}

Similarly, on the HT-CC dataset (Table~\ref{tab:model_comparison_htcc}), we observe a pattern similar to the AMR task: high accuracy models (e.g. MobileNet, MobileViT, CLDNN2) achieve 93\%-95\% accuracy but require up to 300$\times$ more parameters and 900$\times$ more MACs than the proposed model, making them unsuitable for real-time edge computing. Lightweight baselines such as DAE and PET-CGDNN exhibit limited accuracy (79\% - 85\%) and also demonstrate unstable accuracy vs. SNRs, as shown in Fig.~\ref{fig:curve_htcc}. The proposed model's hyperparameters were designed for the AMR tasks but only minimal CPU-side tuning is required to increase the accuracy by adapting to the HT-CC spectral characteristics: increasing the attention bottleneck width (reduction ratio from 9 to 3) and modestly widening the first dense layer (32 to 48 neurons). This yields a task-tuned variant with 90.1\% accuracy, and one to three orders of magnitude lighter than all high-accuracy baselines. These adjustments affect only CPU-executed layers, validating the benefit of our heterogeneous design: task adaptation without hardware changes.

\begin{table}[t]
    \setlength{\tabcolsep}{6pt}
    \centering
    \caption{Ablation study on RadioML2018 dataset. The first row corresponds to the baseline model in~\cite{emad}.}
    \begin{tabular}{cccccc}
    \hline
    \textbf{Decimation} 
    & \textbf{Attn} 
    & \textbf{Redu.} 
    & \textbf{\#Param.} 
    & \textbf{kMACs} 
    & \textbf{Acc.} \\ 
    \hline
    --    & -- & -- 
        & 295,055 & 3,648 & 70.1\% \\
    --    & -- & \checkmark 
        & 148,688 & 2,855 & 69.9\% \\
    --    & \checkmark & \checkmark 
        & 148,976 & 2,965 & 71.5\% \\
    LSDec & -- & -- 
        & 37,016  & 424   & 71.9\% \\
    LSDec & -- & \checkmark 
        & 19,673  & 333   & 71.7\% \\
    Fixed & \checkmark & \checkmark 
        & 19,952  & 340   & 64.2\% \\
    \rowcolor{gray!20} LSDec & \checkmark & \checkmark 
        & 19,961  & 346   & 73.3\% \\
    \hline
    \end{tabular}
    \label{tab:comp_rml2018}
    \vspace{0.15cm}
    \footnotesize
    \\
    Fixed: Traditional fixed-rate downsampling without learned weights. LSDec: Learnable streaming decimator. Attn: Attention module. Redu.: Filters reduction of Conv2d and max-pooling after the second Conv1d.
    \vspace{-0.8em}
\end{table}


\subsubsection{GNSS Jamming}

For GNSS jamming classification (Table \ref{tab:model_comparison_gps}), the dataset provides a single controlled SNR condition; cross-SNR generalization is outside the dataset scope. The proposed model attains a $>$ 99\% accuracy, comparable to state-of-the-art architectures but with five to six significantly lower computational complexity and thousands of times fewer parameters than the largest baselines. Since GNSS and AMR are both critical to UAV communication and navigation~\cite{uavamr,gnssuav}, the extremely small footprint ($\sim$20k parameters, 374 kMACs) makes the model especially suitable for power-efficient and latency-critical edge computing on UAVs.

\subsubsection{Differences between models and datasets}

The three datasets contain the information the neural network is looking for (modulation scheme, covert channel, jamming) embedded in different fields of the transmission frame. For example, in the HT-CC dataset, there are four different underlying HT designs, with two leaking the information into the preamble, one in the payload data, and one across the entire frame. In AMR and GNSS jamming, the information is found primarily in the payload data field. The frame differences vary in each task, e.g., with respect to nominal for HT-CC and GNSS jamming, across underlying HT for HT-CC, and among different modulation schemes for AMR. These structural differences influence model performance (baseline models and the proposed one), as different models are optimized to capture different types of patterns. Recurrent models such as CLDNN2 are better suited for tracking long-range temporal dependencies, whereas image recognition CNN-based approaches excel at identifying spatial patterns within the on-the-fly image-encoded raw I/Q feature representations.


\subsection{Ablation Study}\label{subsec:ablation_study}

We perform an ablation analysis on the RadioML2018 dataset, the most challenging and diverse of the three, allowing architectural differences to be clearly exposed. Table \ref{tab:comp_rml2018} summarizes the impact of each component: LSDec, the attention module, and the filter-reduction/fused-pooling configuration. The first row of Table~\ref{tab:comp_rml2018} corresponds to the baseline CNN from~\cite{emad}, which operates directly on the full length of the raw I/Q sequence without any decimation, yielding 295k parameters and 3,648~kMACs. With a significantly shorter feature vector after flattening, the first dense layer has significantly lower fully-connected weights. We also have a conspicuous computational complexity reduction (-88\% kMACs) as the parallel convolutional filters operate on the notably shorter decimated feature map sequence, which was produced by the LSDec that reduces the sequence length by a factor of $D$=8 (from 1024 to 128) before the first Conv layer. Simple fixed decimation without learned weights would achieve the same parameter reductions, but at the cost of discarding critical signal information, causing accuracy to plummet to 64.2\%. LSDec avoids this by performing feature extraction simultaneously with decimation using a trained kernel that learns to preserve discriminative I/Q.

LSDec provides the largest hardware benefit with decimation, reducing parameters from 295k → 37k (–87.5\%) and computational complexity from 3648 → 424 kMACs (–88.4\%). This is primarily driven by the reduction in spatial sequence length (e.g. 1024$\to$128) before the subsequent layers. This also improves accuracy from 70.1\% to 71.9\% with trained convolutional feature extraction capabilities: hence the name learnable decimation. This shows that learned temporal compression is more effective than fixed decimation and aligns naturally with the accelerator’s streaming data path.

Filter reduction and fused pooling further shrink the model to 19.7k parameters with only a 0.2\% accuracy drop, making it possible for resource-limited FPGA deployments.

The attention block yields accuracy gains both with and without LSDec: accuracy increases to 73.3\% with LSDec, while the full model still achieves a 93.2\% parameter reduction relative to the baseline. This indicates that attention effectively recovers discriminative capacity after aggressive compression, enabling a compact but high-performing model.

Overall, the ablation results validate the co-design philosophy: LSDec enables substantial parameter reduction, filter-reduction and pooling minimizes memory and compute, and attention recovers accuracy, jointly producing a model well suited for real-time, edge-computing accelerators.

\section{Hardware Results}
\label{sec:hw_results}
Unlike prior FPGA-only accelerators that are optimized exclusively for minimum inference latency, the proposed design targets deployment on resource-efficient edge SoCs. Consequently, direct latency comparisons with large FPGA pipelines are not meaningful. Instead, our evaluation emphasizes balanced CPU–accelerator co-execution, power and resource efficiency, and the ability to sustain real-time throughput under realistic SDR front-end constraints for multi-task adaptability.

\subsection{CPU Profiling}
We first profile the full forward pass on the Cortex-A53 CPU by re-implementing the pre-trained model in C using a flat RMO memory layout. An SIMD version is also implemented with ARM Neon intrinsics to accelerate the CPU execution on each core. To establish a worst-case scenario baseline under conservative CPU availability assumptions, this profiling is conducted as a single-thread case study to assess the worst-case performance guarantee. The co-scheduling algorithm automatically rebalances as more cores or faster CPUs are available, using either RAP or static profiling. The code is compiled directly on the CPU for execution. The purely CPU-based execution for the complete AI model requires 1,114 $\mu s$. Layer-level profiling shows that the Conv2d layer requires 353 $\mu s$ and the Conv-MaxPool layer requires 754 $\mu s$. The attention and all dense layers require $<20 \mu s$ combined. These results demonstrate that convolutional layers dominate runtime, motivating their execution on a specialized hardware accelerator. The attention layer is highly efficient on CPU ($<10 \mu s$) and also for task flexibility (e.g. changing the reduction ratio to adapt to another RF task), thus it is completely executed on the CPU directly. 

Our CPU baseline uses direct Neon SIMD with row-major indexing rather than standard im2col+GEMM libraries that lack native support for our custom W8A16-INT quantization for RF tasks, and widening weights to 16 bits requires an extra memory pass. Using the Conv2d layer as a case study: the INT16 input of shape (1,2,128) and kernel size (2$\times$8) with 36 filters produces a 16 $\times$ 121 INT16 (3,872 bytes) workspace. This 7.5$\times$ memory expansion over the 512-byte input eliminate cache-tiling benefits. For the Conv-MaxPool layer, the required im2col workspace (49KB) exceeds the Cortex-A53's 32KB L1 data cache, introducing cache-refill overheads that outweigh GEMM efficiencies. Direct SIMD avoids this data duplication, representing a highly optimized embedded baseline. Regarding the SIMD execution, vectorization is applied across the inner MAC loop. The smaller kernels require gathered access, frequent address updates, and horizontal reductions. This makes the convolution path memory-bound, keeping the practical throughput below the nominal 8$\times$ lane width. In contrast, the fully-connected layers operate on continuous and cache-resident vectors with simple strides and deep inner loops, allowing near-ideal vector issue and accumulation. 

\subsection{Hardware Accelerator: Latency and Resource }

\begin{figure}[t]
\centerline{\includegraphics[width=\linewidth]{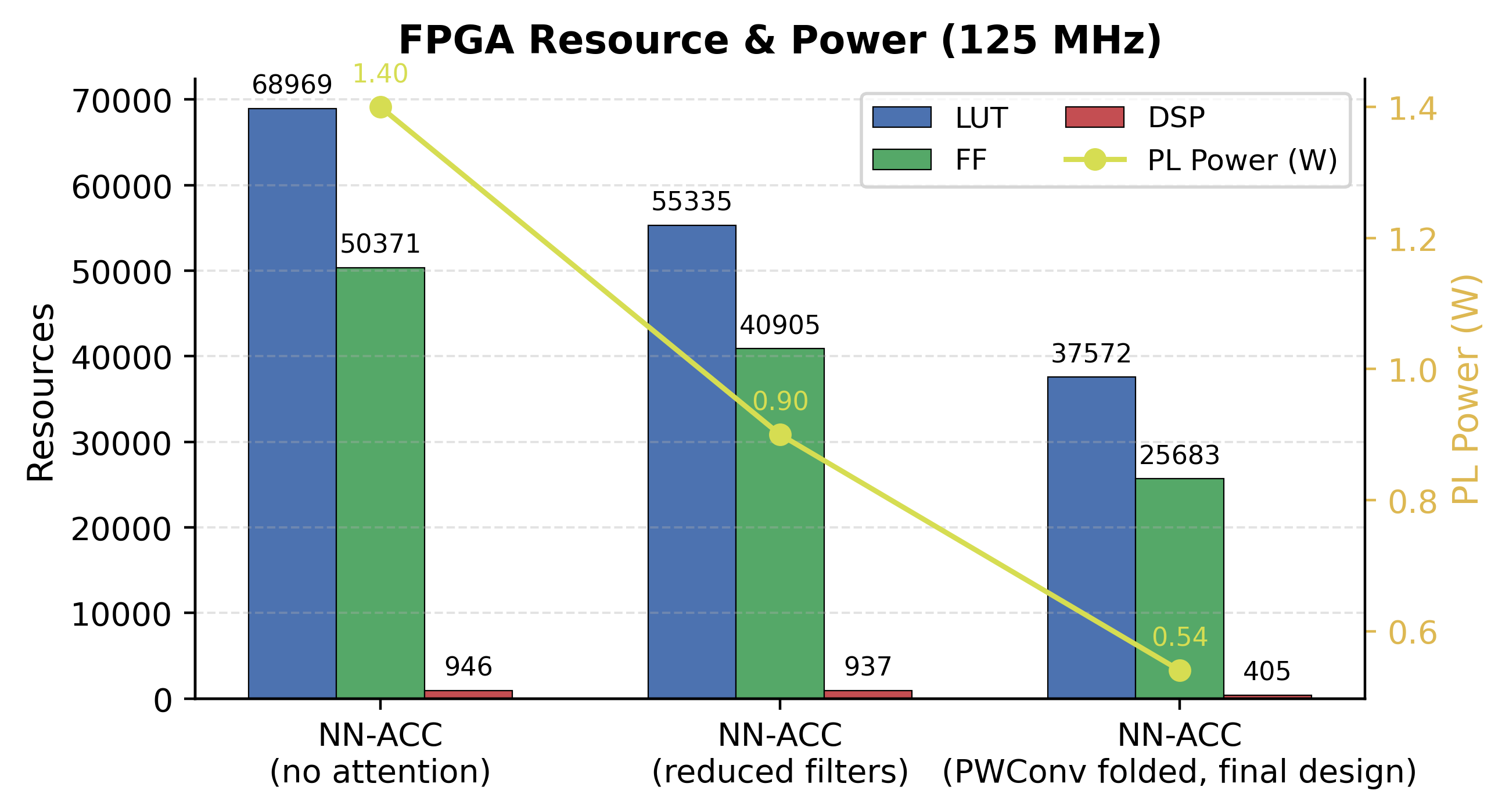}}\vspace{-0.4cm}
\caption{FPGA resource and power comparison at 125\,MHz. Scenario 1: accelerator with AXI interfaces, attention offloaded. Scenario 2: heterogeneous split with reduced accelerator filters. Scenario 3 (final): PWConv folding to reuse conv engines, cutting resource usage and power. Only FPGA PL power is reported to ensure fair comparison with existing work, with the assumption that an embedded CPU is always present in a system.}
\label{fig:resource}\vspace{-0.4cm}
\end{figure}

The proposed accelerator implements deeply pipelined convolution engines with input–weight forwarding and a dual-pipeline fused Conv–MaxPool unit (Fig. \ref{fig:soc}). Mapping all convolutional filters to the NN accelerator on FPGA PL reduces both the Conv2d and Conv–MaxPool layers to 42 $\mu s$ each, and accelerates the first dense layer from 17 $\mu s$ to 7 $\mu s$. However, this configuration consumes 69k LUTs, 50k FFs, and 946 DSP slices, with 1.4 W dynamic power at 125 MHz. To achieve deployability on power- and area-constrained SoCs, we introduce the heterogeneous execution model in which a subset of convolutional filters and dense neurons is intentionally offloaded on the CPU. This reduces resource usage to 55k LUTs, 41k FFs, and 937 DSPs (0.9 W). After applying PWConv folding, the final accelerator on FPGA PL uses only 38k LUTs, 26k FFs, and 405 DSPs, consuming 0.54 W while sustaining 98 $\mu s$ end-to-end latency, which is measured on the MPSoC ZCU104 board and includes all synchronization, cache flush, and DMA setup overhead, as illustrated in Fig. \ref{fig:resource}. Compared to state-of-the-art (previously seen in Table~\ref{tab:fpga_related}), the current design adds the lowest power (0.54W), lowest FPGA resource usage, well below prior art while achieving satisfying inference speed and striking the power, performance, area (PPA) balance.

\begin{figure}[t]
\centerline{\includegraphics[width=\linewidth]{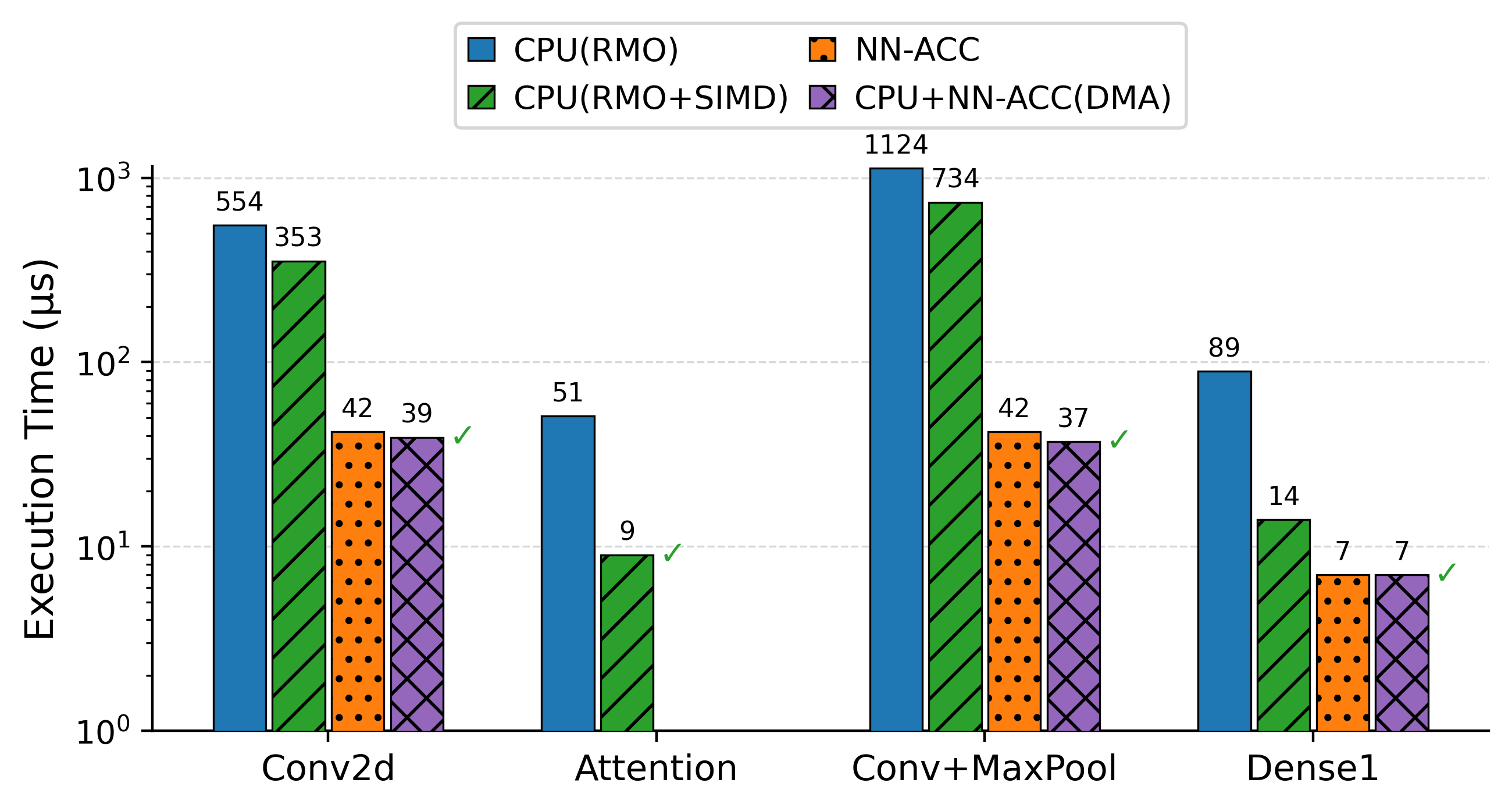}}\vspace{-0.4cm}
\caption{Execution time profiling. The last dense layer only requires 1 $\mu s$ and is discarded on the figure. Log-scale is applied for visual clarity due to huge execution time difference.
}
\label{fig:profiling}\vspace{-0.6cm}
\end{figure}
\begin{table}[t]
\centering
\caption{Comparison with Recent state-of-the-art Sub-Watt(PL) FPGA Implementations}
\begin{tabular}{lccc}
\hline
\textbf{Metric} & \textbf{TWC'25~\cite{MRTransformer}} & \textbf{TCASAI'25~\cite{tcasai}} & \textbf{This Work} \\
\hline
Weights Quant.     & 16-bit          & 16-bit          & \textbf{8-bit} \\
LUT              & 43{,}380        & 83{,}572        & \textbf{37{,}572} \\
FF               & 29{,}457        & 45{,}060        & \textbf{25{,}683} \\
DSP              & 328             & 297             & \textbf{405} \\
Power            & 0.60 W          & 0.36 W          & \textbf{0.54 W} \\
Latency          & 2 ms            & 453.1 $\mu$s    & \textbf{98 $\mu$s} \\
\hline
\end{tabular}
\label{tab:fpga_sota}\vspace{-0.4cm}
\end{table}

Regarding the Conv-MaxPool fusion: the second convolution layer produces a (9,1,116) INT16 intermediate tensor (2,088 bytes) and implementing separate Conv and MaxPool IPs would require a DDR roundtrip and extra DMA transaction. The fused dual-pipeline engine fuses two operators and halves memory traffic simultaneously. The Conv-MaxPool layer can match the Conv2d engine latency, producing a more balanced inference pipeline. The trade-off is reduced flexibility: the fused engine assumes max-pooling always follows this convolution layer with a maxpooling with a stride of two. Nevertheless, this is appropriate for the co-designed Attention-CNN model which uses this architecture across all three RF tasks and achieves good accuracy and satisfies design goals.

Execution timing for all configurations: CPU-only, CPU-SIMD, full NN accelerator (NN-ACC), and heterogeneous accelerator (reduced filters and includes all AXI data interfaces, memory controller, and DMA), is shown in Fig. \ref{fig:profiling}. As expected, convolution benefits most from hardware parallelism, while attention and dense layer operations remain efficient on the CPU. For reference, the A53 cluster consumes 2.64 W dynamic power. 

In practical IoT and embedded platforms, a CPU is always present to manage control flow, scheduling, and system tasks; therefore, some level of CPU activity is unavoidable regardless of the accelerator design. Reporting accelerator-only power can thus be misleading, since real deployments inherently incur CPU dynamic power whenever computation is orchestrated. Our heterogeneous design follows this system model by assigning lightweight operations to the CPU while offloading compute-intensive layers to the FPGA, enabling an efficient division of labor. Compared with recent state-of-the-art sub-W PL implementations (see Table~\ref{tab:fpga_sota}), our approach achieves the best trade-off: although certain designs such as SNNs achieve low power, they typically exhibit limited interoperability with CPU-side processing and require significant LUT/FF resources whereas transformer models introduce significant latency. In contrast, our design maintains low PL power, reduces resource usage, and achieves lower latency, offering a more practical and balanced solution for embedded inference.

\begin{figure}[t]
\centerline{\includegraphics[width=\linewidth]{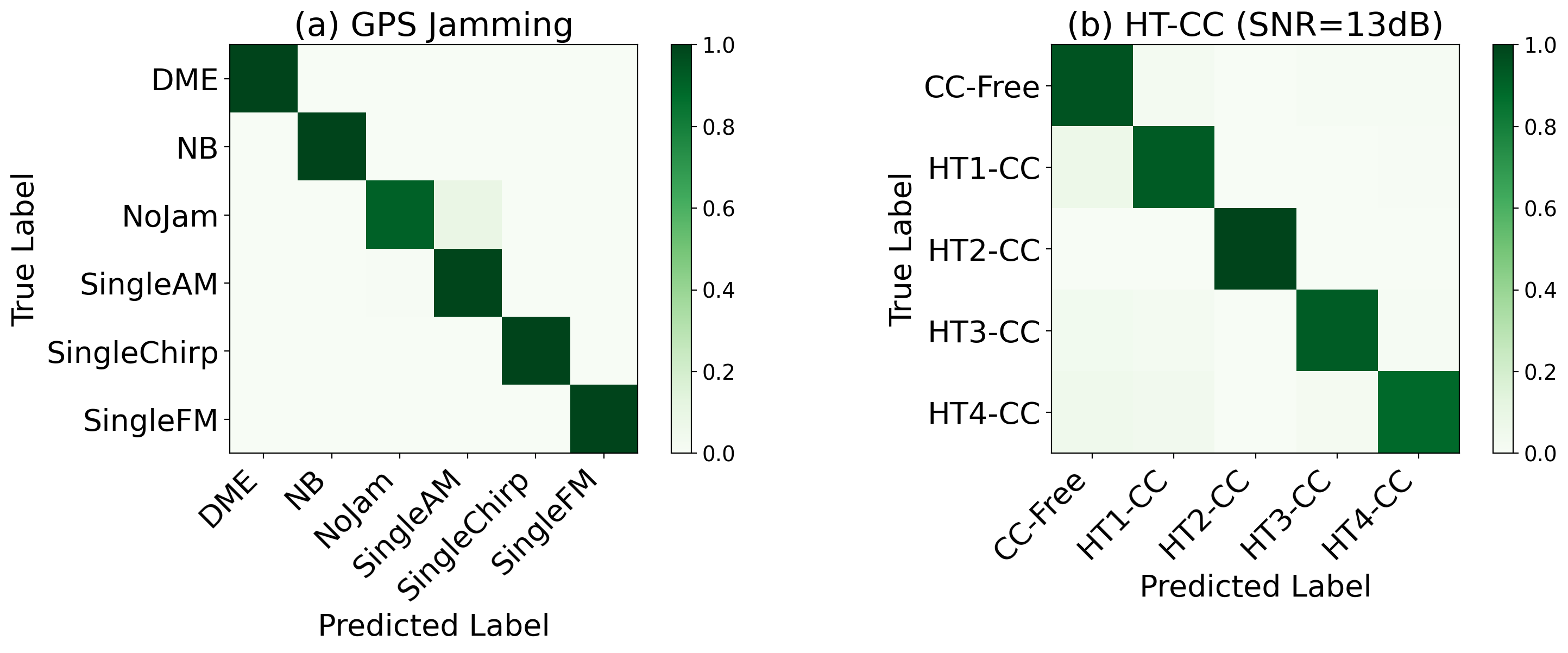}}\vspace{-0.4cm}
\caption{Confusion matrix on GNSS jamming and HT-CC (SNR=13~dB) detection tasks with the quantized model.}
\label{fig:cm_gps_htcc}\vspace{-0.4cm}
\end{figure}

\subsection{Co-Scheduling Algorithm Case Study on Conv2d Layer}
\label{sec:caseco}
To illustrate the operation of the co-scheduling algorithm~\ref{alg:cosched} assuming no padding, we present a case study of the first Conv2d layer. Empirically measured execution times are $T_a = 42\mu s$ for the accelerator and $T_c=353 \ \mu s$ for the CPU (SIMD). The unconstrained optimal filter split, determined by the standard load-balancing condition, is approximately 33 filters on the accelerator and 3 on the CPU. However, correctness requires cache-line-aligned writes due to the shared contiguous memory buffer. The output tensor has shape $36 \times 121$ with 16-bit precision after requantization, corresponding to $121\times 2 = 242$ bytes per channel. Since we have 64-byte cache lines, 242 $\times$ 33 = 7,986 bytes is not aligned (7986 mod 64 =34). The co-scheduling algorithm therefore selects the closest aligned partition: 32 filters on the accelerator and 4 on the CPU. This preserves throughput, prevents coherence stalls, and ensures deterministic correctness. While this case study establishes our worst-case single-thread baseline, real-world deployment adapts to dynamic CPU availability. As detailed in Section~\ref{sec:algo}, if $M$ cores are allocated, the framework utilizes core-symmetry constraints and memory padding to seamlessly distribute the expanded software partition, further lowering the dynamic power of the accelerator engines.

Following the sequence in Section~\ref{sec:algo}, the worst-case synchronization overhead consists of two components: execution asymmetry, which the co-scheduling algorithm (Alg.~\ref{alg:cosched}) already minimizes, and the cache flush duration itself. The Conv2d layer represents the worst-case synchronization scenario, as its per-filter output (242 bytes) is the largest across layers. For the single-core baseline, the flush covers the CPU-written region: 4 filters $\times$ 242 bytes = 968 bytes (16 cache lines), bounding the peak synchronization latency to approximately 1.3~$\mu$s on the Cortex-A53. This minimal overhead is absorbed within the measured 98~$\mu$s end-to-end latency. With $M$ cores, dirty cache lines are flushed in parallel, ensuring this overhead does not increase.

Finally, compared with conventional processors, ARM CPUs typically consume $\sim$750mW/Core under load, yet remain an order of magnitude slower than GPUs on NN workloads. Conversely, a high-end GPU (NVIDIA A100) achieves 13 $\mu s$ inference but consumes in the range of 250–400 W, making it incompatible with any embedded RF platforms. These comparisons highlight the suitability of the proposed heterogeneous architecture for real-time spectrum intelligence.

\vspace{-0.6em}
\subsection{Quantization for Full-Integer Inference} 

The proposed model trained in PyTorch and the W8A16-INT quantized model (described in Section~\ref{sec:acnnmodel}) are rewritten into a custom inference graph (C and Verilog) designed for full-integer execution. For the AMR task on RadioML2018, the floating-point model achieves 73.3\% average accuracy across all SNRs and 99.6\% for SNR $>$ 0 dB. After quantization, accuracy remains high at 72.2\% and 99.4\%, respectively, with a total memory footprint of only 20~KB. On the HT-CC dataset, quantization yields 89.3\% accuracy versus the 90.1\% floating-point baseline. On the GNSS jamming dataset, accuracy after quantization remains 98.5\% across all classes. Figs.~\ref{fig:curve_rml2018} and~\ref{fig:curve_htcc} confirm that the quantized model deployed on hardware (green curves) closely tracks the floating-point GPU model (blue curves) for all SNRs. Notably, for the AMR task, both models exceed 99\% accuracy from 4 dB onward, indicating that quantization does not affect the SNR regime of practical interest. Representative confusion matrices for the quantized models on the (GNSS jamming dataset and the HT-CC dataset (SNR=13 dB) are shown in Fig.~\ref{fig:cm_gps_htcc}, confirming that class-level performance is preserved.

\vspace{-0.25em}
\section*{Conclusion}

In this work, we presented, to the best of our knowledge, the first heterogeneous AI accelerator designed specifically for multi-task RF signal recognition at the edge. By deploying the learnable streaming decimator directly at the ADC interface, the system bridges the speed mismatch between high-rate RF front-ends and power-efficient AI cores without full-frame buffering. A cache-coherent co-scheduling algorithm ensures safe and balanced parallel execution across CPU and accelerator cores under a unified memory architecture. The proposed heterogeneous framework provides a highly practical, scalable solution for real-time, multi-task spectrum intelligence on low-power edge devices, achieving an end-to-end inference latency of 98 $\mu$s. The resulting model achieves the best accuracy-complexity trade-off among existing RF classifiers while supporting rapid task tuning by modifying CPU-executed parts of the model, demonstrated across AMR, HT-CC, and GNSS-jamming classification, eliminating any hardware redesign. This flexibility, combined with sub-W power consumption and real-time throughput, makes the proposed accelerator well-suited for embedded SDR platforms and edge-computing systems such as UAVs. Future extensions include scaling for additional RF tasks and exploring other models for further optimized accuracy within the same heterogeneous paradigm.


\bibliographystyle{ieeetr}
\bibliography{ref}


 





\end{document}